\documentclass[12pt]{WileyMSP-template}
\usepackage{subfig}
\usepackage{amsmath}
\usepackage{comment}
\usepackage{xcolor}
\usepackage{siunitx}
\usepackage{upgreek}
\usepackage{mathptmx}
\usepackage{cite}
\usepackage{pdfpages}
\begin{document}
\linespread{1.25}
\pagestyle{fancy}
\newcommand{\Irf}{I_\text{RF}}
\newcommand{\Sa}{S^{\text{AHE/art}}_\text{XY}}
\newcommand{\Sp}{S^{\text{PHE/art}}_\text{XY}}
\newcommand{\Sm}{S^{\text{AMR/art}}_\text{XX}}
\newcommand{\Aa}{A^{\text{AHE}}_\text{XY}}
\newcommand{\Ap}{A^{\text{PHE}}_\text{XY}}
\newcommand{\Am}{A^{\text{AMR}}_\text{XX}}

\title{Separation of Artifacts from Spin-Torque Ferromagnetic Resonance Measurements of Spin-Orbit Torque for the Low-Symmetry van der Waals Semi-Metal ZrTe$_3$}

\maketitle

% Author: Please give full first and last names for authors and include * after the name of all corresponding authors

\author{Thow Min Cham, Saba Karimeddiny, Vishakha Gupta, Joseph A. Mittelstaedt and Daniel C. Ralph$^{*}$}

% Dedication
\dedication{}

% Affiliations: Please provide adacemic titles (Prof. or Dr.) for all authors where applicable, and include an institutional email address for all corresponding authors
\begin{affiliations}
Thow Min Cham, Saba Karimeddiny, Vishakha Gupta, Joseph A. Mittelstaedt, \\ Prof. Daniel C. Ralph\\
Department of Physics, Cornell University, Ithaca, NY 14850, USA\\
Email Address: dcr14@cornell.edu\\
Prof. Daniel C. Ralph\\
Kavli Institute at Cornell, Ithaca, NY 14853, USA\\
\end{affiliations}

% Keywords: Please provide a minimum of three and a maximum of seven keywords, separated by commas


% Abstract should be written in the present tense and impersonal style (i.e., avoid we), and be at most 200 words long
\begin{abstract}

  Spin-orbit torques generated by exfoliated layers of the low-symmetry semi-metal ZrTe$_3$  are measured using the spin-torque ferromagnetic resonance (ST-FMR) technique. When the ZrTe$_3$ has a thickness greater than about 10 nm, artifacts due to spin pumping and/or resonant heating can cause the standard ST-FMR analysis to overestimate the true magnitude of the torque efficiency by as much as a factor of 30, and to indicate incorrectly that the spin-orbit torque depends strongly on the ZrTe$_3$ layer thickness. Artifact-free measurements can still be achieved over a substantial thickness range by the method developed recently to detect ST-FMR signals in the Hall geometry as well as the longitudinal geometry. ZrTe$_3$/Permalloy samples generate a conventional in-plane anti-damping spin torque efficiency $\xi_{||}^{\text{DL}}$ = 0.014 $\pm$ 0.004, and an unconventional in-plane field-like torque efficiency $|\xi_{||}^{\text{FL}}|$ = 0.003 $\pm$ 0.001.  The out-of-plane anti-damping torque is negligible. We suggest that artifacts similarly interfere with the standard ST-FMR analysis for other van der Waals samples thicker than about 10 nm.

\end{abstract}

\section{Introduction}
Spin-orbit torques can provide efficient switching of magnetization in nanoscale magnetic \\devices.\textsuperscript{\cite{Brataas2012, Wang2013, Manchon2019}} The torques generated by the heavy metals Pt,\textsuperscript{\cite{Miron2010, Miron2011, Liu2011, Zhang2015, Pai2015, Qiu2016, Zhao2019}} Ta,\textsuperscript{\cite{Liu2012, Yu2014, Cubukcu2018}} W\textsuperscript{\cite{Pai2012, Demasius2016}} and their alloys\textsuperscript{\cite{Zhu2019, Cha2020}} have been studied in detail.  
More recently, spin-orbit torques generated by van der Waals materials have been an increasing focus, in order to study the consequences of spin-momentum coupling in topological materials and to generate spin-orbit torques with unconventional orientation using low-symmetry materials.  
Recent experiments in these categories include studies of transition metal dichalcogenides (TMDs) MoS$_2$,\textsuperscript{\cite{Zhang2016, Cheng2016, Safeer2019}} WTe$_2$,\textsuperscript{\cite{MacNeill2017_STFMR, Macneill2017b, Li2018, Shi2019}} WS$_2$,\textsuperscript{\cite{Lv2018}} NbSe$_2$,\textsuperscript{\cite{Guimaraes2018}}  
TaTe$_2$,\textsuperscript{\cite{Stiehl2019b}}
MoTe$_2$,\textsuperscript{\cite{Stiehl2019, Song2020, Liang2020}}  PtTe$_2$,\textsuperscript{\cite{Xu2020}} TaSe$_2$,\textsuperscript{\cite{Husain2020}} WSe$_2$,\textsuperscript{\cite{Novakov2021, Hidding2021}} and Cd$_3$As$_2$,\textsuperscript{\cite{Yanez2021}} as well as studies of the topological insulators Bi$_2$Se$_3$,\textsuperscript{\cite{Mellnik2014, Wang2015, Dc2018}} BiSb\textsuperscript{\cite{Khang2018}} and magnetically-doped BiSbTe.\textsuperscript{\cite{Fan2014, Fan2016}}

\vspace{0.5cm}
When devices are made from exfoliated samples of van der Waals materials, it can sometimes be difficult to isolate layers thinner than a few 10's of nanometers, so the spin-orbit layers can be much thicker than for typical measurements of heavy metals, where the spin-orbit layers are generally much thinner than 10 nm.  Here we illustrate using ZrTe$_3$/Permalloy bilayers that extra care is required when employing the spin-torque ferromagnetic resonance (ST-FMR) technique\textsuperscript{\cite{Liu2011, Tulapurkar2005, Sankey2006, Fang2011}} in devices with thicker spin-orbit layers, because the magnitude of artifacts due to spin pumping\textsuperscript{\cite{Tserkovnyak2002, Tserkovnyak2002b, Mosendz2010}} and resonant heating\textsuperscript{\cite{Holanda2017}} grow relative to the spin-orbit-torque signals as a function of increasing layer thickness.  One signature of such artifacts is an apparent dependence of the spin-torque efficiency on the spin-orbit layer thickness for layers much thicker than a typical spin diffusion length. We demonstrate that a recently-introduced modification of the ST-FMR technique,\textsuperscript{\cite{Karimeddiny2020}} in which the ST-FMR signals are measured in the Hall geometry as well as the usual longitudinal geometry,\textsuperscript{\cite{Bose2017, Kumar2017}} allows more-accurate measurements of the spin-orbit torques, separated from artifacts due to spin pumping and resonant heating, without any significant added experimental effort.

\begin{figure}[htpb]
    \centering
    \subfloat{
    \begin{minipage}{0.5\textwidth}
        \centering
        \includegraphics[width=0.6\textwidth]{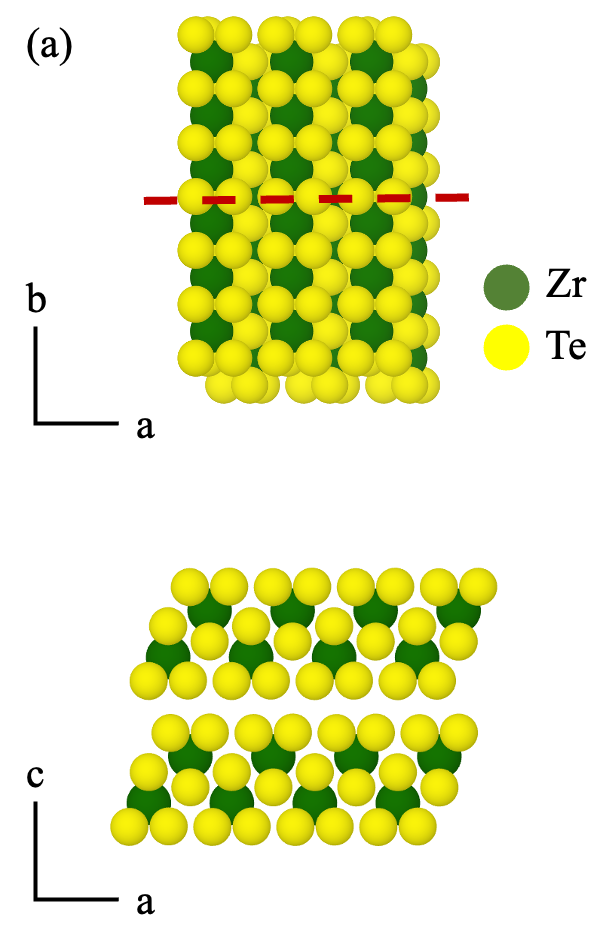}
    \end{minipage}\hfill}
    \subfloat{
    \begin{minipage}{0.5\textwidth}
        \centering
        \includegraphics[width=1.0\textwidth]{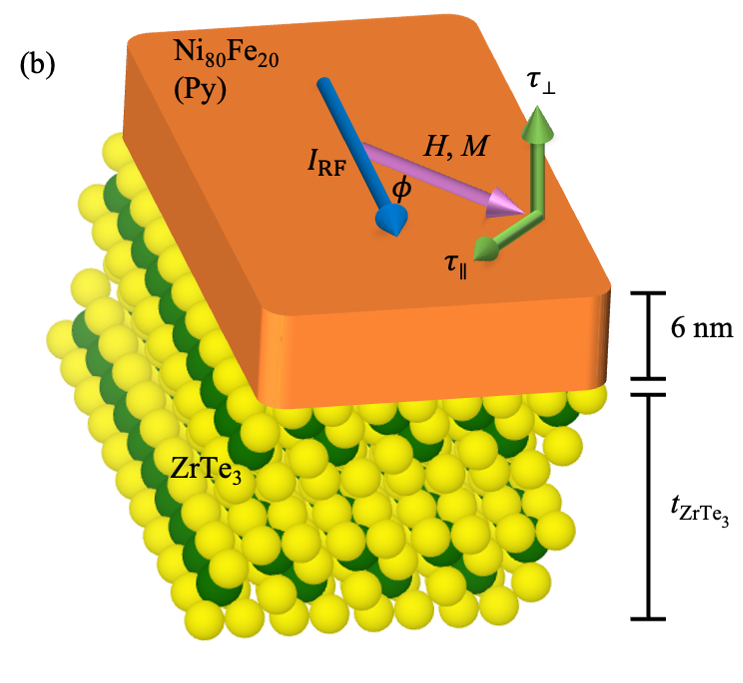}
    \end{minipage}\hfill}
    \caption{a) Crystal structure of ZrTe$_3$ in the a-b and c-a planes. The red dashed line indicates the intersection of the mirror plane with the a-b plane. b) Schematic of a ZrTe$_3$/Permalloy(Py) heterostructure with applied current flow along the b axis (perpendicular to the mirror plane), a geometry for which unconventional spin-orbit torques are symmetry-allowed.\cite{Momma:db5098}}
    \label{fig:ZrTe3Xtal}
\end{figure}

\begin{figure}[htpb]
    \centering
    \subfloat{
    \begin{minipage}{0.5\textwidth}
        \centering
        \includegraphics[width=0.8\textwidth]{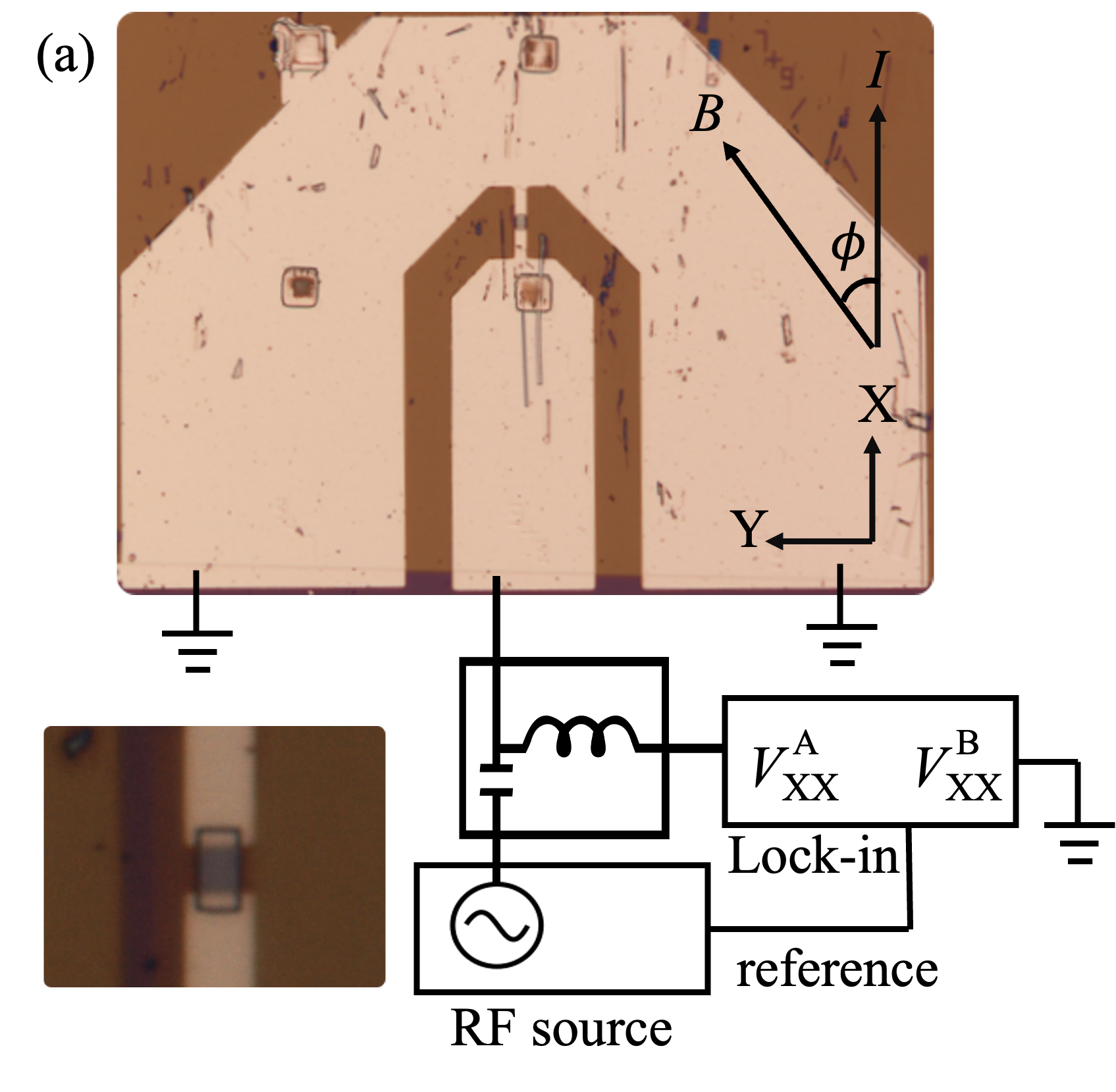}
    \end{minipage}\hfill}
    \subfloat{
    \begin{minipage}{0.5\textwidth}
        \centering
        \includegraphics[width=0.8\textwidth]{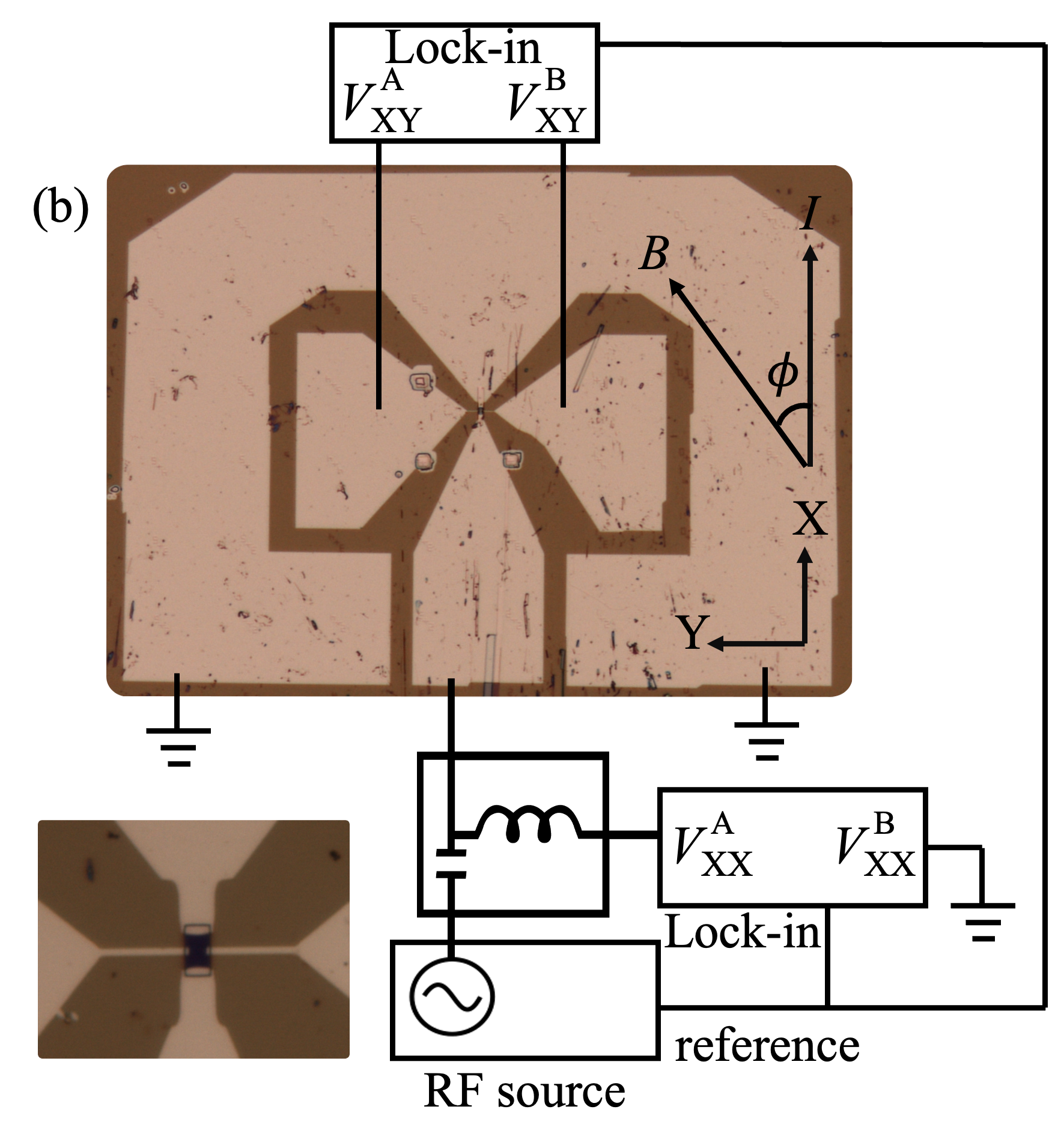}
    \end{minipage}\hfill}
    \caption{Schematic of the device geometry and measurement circuit for (a) conventional ST-FMR and (b) Hall-detected ST-FMR. For the Hall-detected ST-FMR measurements, an additional lock-in amplifier connected across the Hall leads is used to read out the transverse mixing voltage $V_\text{XY}$ = $V^\text{A}_\text{XY}$ - $V^\text{B}_\text{XY}$.}
    \label{fig:schematic}
\end{figure}

\vspace{0.5cm}
ZrTe$_3$ has the space group P2$_1$/m (no.\ 11), with a screw axis along the Zr chain and a mirror plane perpendicular to the screw axis\textsuperscript{\cite{Geremew2018, Liu2020}} as shown in \textbf{Figure\ \ref{fig:ZrTe3Xtal}}a. ZrTe$_3$ flakes exfoliate into narrow nano-ribbons extended along the b axis, perpendicular to the a-c mirror plane, with typical dimensions of \SI{20}{\micro\meter} x \SI{5}{\micro\meter}. The crystal axis of the exfoliated flakes can be easily identified by the orientation of the nano-ribbons. We prepare spin-orbit-torque test structures of ZrTe$_3$ with varying thicknesses $t\rm_{ZrTe_3}$ capped with 6 nm of Permalloy (Py, Ni$_{80}$Fe$_{20}$). To avoid air exposure of the ZrTe$_3$, we perform the final step of exfoliation in the load-lock chamber of our sputter system, at pressures $<$ 10$^{-6}$ torr. We then deposit the Py via magnetron sputtering without breaking vacuum. The Py has in-plane magnetic anisotropy. Figure \ref{fig:ZrTe3Xtal}b shows a schematic of the Py-covered ZrTe$_3$ heterostructure. We have studied two types of device geometries made by electron-beam lithography and ion milling (see the Experimental Section): standard ST-FMR bars in which only longitudinal electrical signals can be measured (\textbf{Figure \ref{fig:schematic}}a) and ST-FMR devices which also contain Hall contacts (with \SI{1}{\micro\meter} side wires) (Figure \ref{fig:schematic}b). The devices with Hall contacts allow a separation of  spin-orbit-torque signals from artifacts due to spin pumping and resonant heating.

\section{Results}
For our ST-FMR measurements, we use a microwave-frequency (9-12 GHz) current source to generate current-induced torques on the Py magnetic layer while sweeping an in-plane magnetic field $B$ at a fixed angle $\phi$ (Figure 1b). When the magnetic field is swept through the resonance condition, the resulting magnetic precession produces resistance oscillations that mix with the applied current to create a DC voltage signal. In addition to this mixing signal, resonant DC voltages that we will describe as artifact voltages can also be generated by (i) spin pumping of spin current from the magnetic layer to the spin-orbit layer together with the inverse spin Hall effect in the spin-orbit layer, (ii) resonant heating that generates a spin Seebeck-induced spin current into the spin-orbit layer together with the inverse spin Hall effect, and (iii) resonant heating that generates a thermal gradient within the conducting magnetic layer together with the Nernst effect. In all cases, these artifacts result in an electric field at resonance that is perpendicular to the magnetization axis, so that the artifacts produce voltage signals sharing same dependence on the angle $\phi$.
We modulate the amplitude of the microwave-frequency current source at 1 kHz and detect the resonant voltage from the sample using a bias tee and lock-in amplifier.  

\vspace{0.5cm}
We will analyze the results from our ST-FMR measurements first using an (incorrect) ``standard'' analysis framework that neglects the contributions from spin pumping and resonant heating.  We will then demonstrate the framework that considers both the longitudinal and Hall-detected resonance voltages, and that allows the artifact signals to be separated from the mixing signal due to the spin-orbit torques.

\subsection{Analysis neglecting artifacts}
\begin{figure}[htpb]
    \centering
    \subfloat{
    \begin{minipage}{0.5\textwidth}
        \centering
        \includegraphics[width=\textwidth]{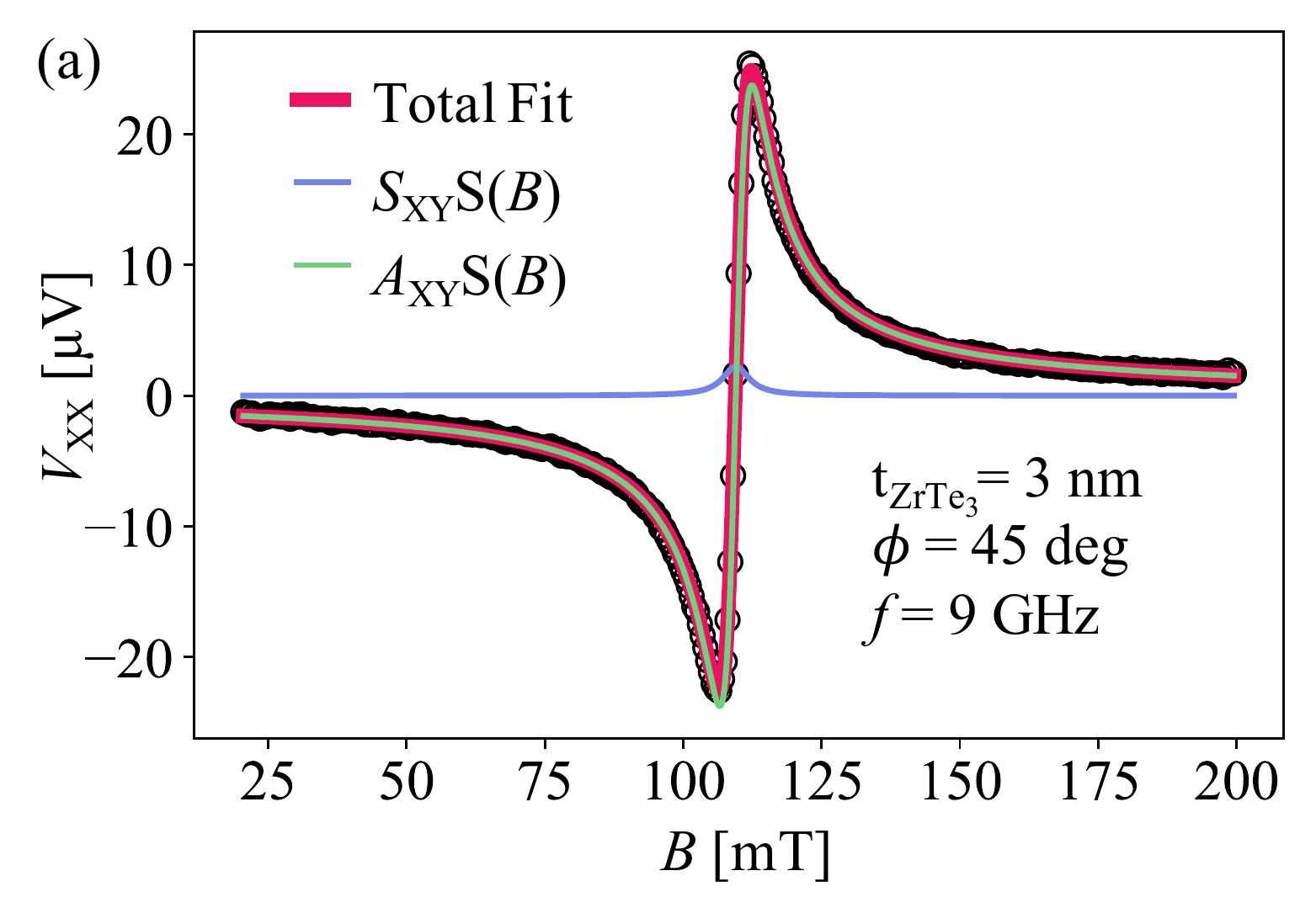}
    \end{minipage}\hfill}
    \subfloat{
    \begin{minipage}{0.5\textwidth}
        \centering
        \includegraphics[width=\textwidth]{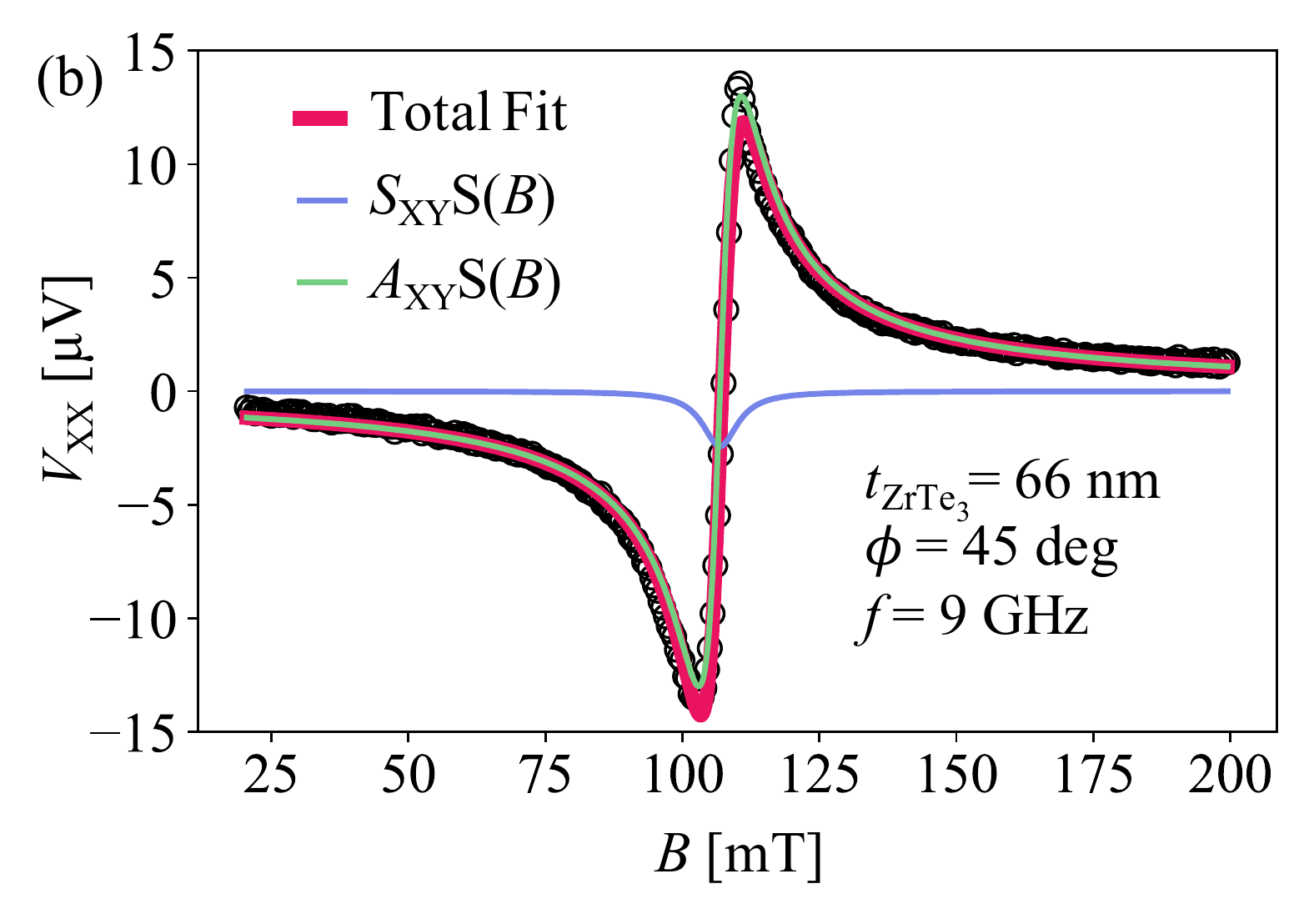}
    \end{minipage}}
    \caption{Longitudinal ST-FMR signal as a function of applied magnetic field at a fixed angle of 45$^{\circ}$ for (a) a 3 nm and (b) a 66 nm thick ZrTe$_3$ flake, along with fits to the sum of antisymmetric and symmetric Lorentzian line shapes.}
    \label{fig:AMR-ST-FMR-res}
\end{figure}

Within the standard ST-FMR analysis framework, only the longitudinal voltage $V\rm_{XX}$ parallel to the applied current is detected.  We fit this resonant voltage to a sum of symmetric and anti-symmetric Lorentzian components, plus a linear background to account for the ordinary Nernst effect (ONE)
\begin{equation}
V_{\text{XX}}(\phi) = S_{\text{XX}}(\phi)\text{S}(B) + A_{\text{XX}}(\phi)\text{A}(B) + V_{\text{ONE}}(\phi)B.
\label{eqn:Vmix}
\end{equation}
Here we define the symmetric and anti-symmetric Lorentzians as S($B$) = $\frac{\Delta^2}{(B-B_0)^2 + \Delta^2}$ and A($B$) = $\frac{\Delta(B-B_0)}{(B-B_0)^2 + \Delta^2}$ where $B_0$ is the resonant field and $\Delta$ is the linewidth. Fits to the data taken at $f$ = 9 GHz and $\phi$ = 45 deg in \textbf{Figure \ref{fig:AMR-ST-FMR-res}}a,b, show good agreement for both thin (3 nm) and thick (66 nm) ZrTe$_3$. 
These data can be analyzed within the framework of the Landau-Lifshitz-Gilbert-Slonczewski equation,\textsuperscript{\cite{SLONCZEWSKI1996L1}} assuming a magnetic layer with in-plane anisotropy and that macrospin magnetic dynamics are a good approximation\textsuperscript{\cite{Tannous_2008}}
\begin{equation}
\frac{\text{d} \hat{m}}{\text{d}t} = - \upgamma \hat{m} \times \vec{B}_{\text{eff}} + \alpha \hat{m} \times \frac{\text{d} \hat{m}}{\text{d}t} + \vec{\tau}_{||} + \vec{\tau}_{\perp},
\label{eqn:LLGS}
\end{equation}
where $\hat{m}$ is the orientation of the magnetic moment, $B_{\text{eff}}$ is an effective magnetic field, $\upgamma$ is the gyromagnetic ratio, $\alpha$ is the Gilbert damping parameter ($\alpha = \upgamma\Delta / \omega$ where $\omega$ is the resonance angular frequency), and $\vec{\tau}_{||}$ and $\vec{\tau}_{\perp}$ are the in-plane and out-of-plane current-induced torques per unit magnetic moment.
If one assumes that the signal is entirely due to the mixing voltage, with no contribution from the artifact voltages, then the amplitudes of the Lorentzian components $S\rm_{XX}$ and $A\rm_{XX}$ depend on the strengths of $\vec{\tau}_{||}$ and $\vec{\tau}_{\perp}$\textsuperscript{\cite{MacNeill2017_STFMR}}
\begin{equation}
S_\text{XX} = \frac{\Irf R_\text{AMR}\sin(2\phi)}{2\alpha \omega^+} \tau_{||}(\phi)
\label{eqn:Vxxs}
\end{equation}
\begin{equation}
A_\text{XX} = \frac{\Irf R_\text{AR}\sin(2\phi)}{2\alpha \omega^+} \frac{\omega_2}{\omega}\tau_{\perp}(\phi)
\label{eqn:Vxxa}
\end{equation}
 where $I\rm_{RF}$ is the applied GHz current and the change in the resistance of the bilayer as a function of in-plane magnetization orientation is $R(\phi)= R_{\text{AMR}}$ cos$^2(\phi)$. We define $\omega_1$ = $\upgamma B_0$, $\omega_2$ = $\upgamma(B_0$ + $\upmu_0M_\text{eff})$, $\omega = \sqrt{\omega_1 \omega_2}$, and $\omega^+$ = $\omega_1$ + $\omega_2$. $B_\text{0}$ is the resonant field and $\upmu_0 M_\text{eff}$ is the perpendicular anisotropy field (positive for a film with in-plane anisotropy).

\begin{figure}[htpb]
    \centering
    \includegraphics[width=\textwidth]{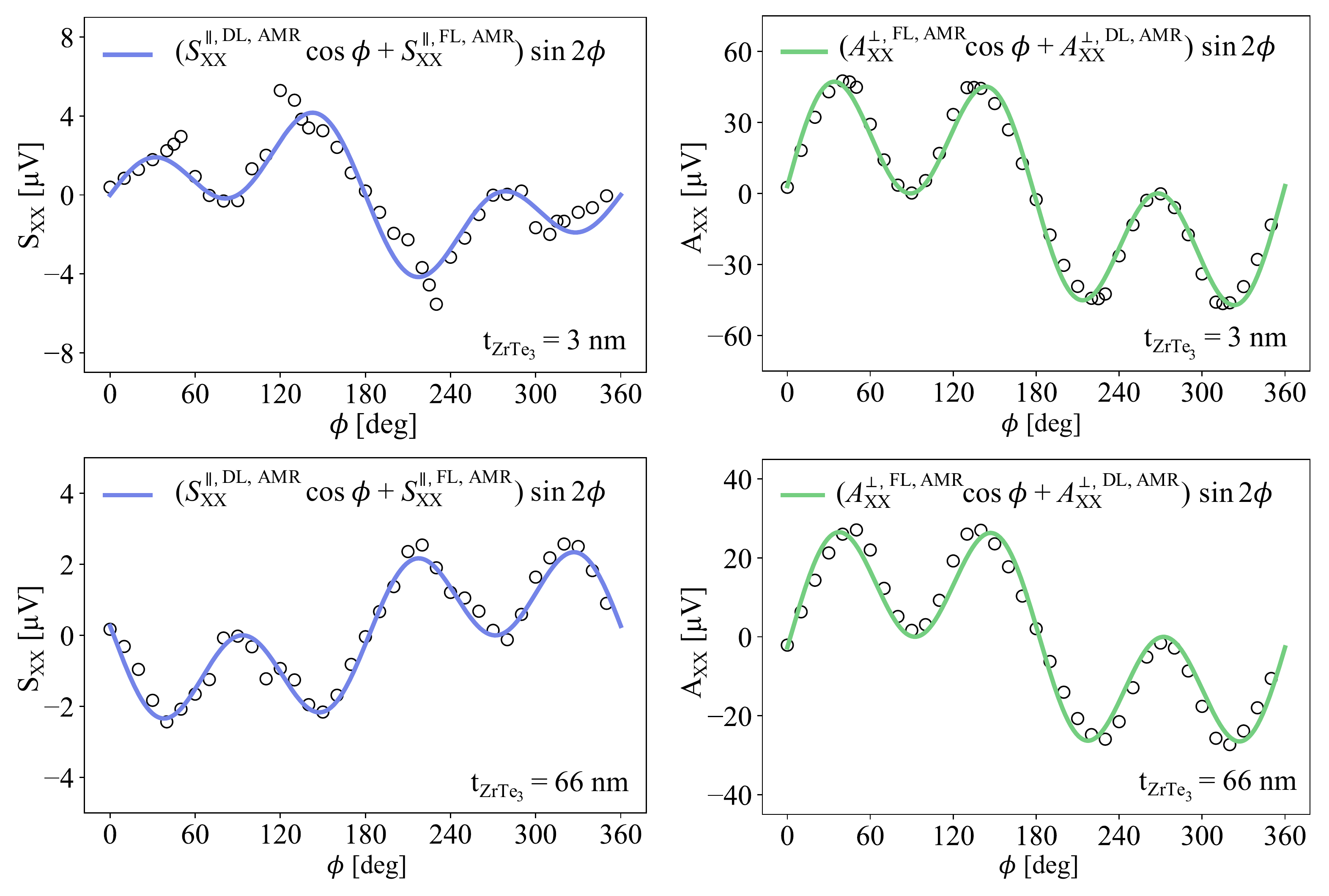}
    \caption{Symmetric ($S_{XX}$) and antisymmetric ($A_{XX}$) components of  longitudinal ST-FMR signals as a function of in-plane  magnetic-field angle for 3 nm and 66 nm thick ZrTe$_3$ flakes, along with fits to Equations (\ref{eqn:VphiS}) and (\ref{eqn:VphiA}).}
    \label{AMR-ST-FMR-SA}
\end{figure}

\vspace{0.5cm}
For high-symmetry spin-orbit materials, the in-plane torque from the spin current $\hat{\sigma}$, is constrained by symmetry to have the anti-damping form $\Vec{\tau_{||}} =  \tau^0_{||} \hat{m}\ \times$ $(\hat{\sigma}$ $\times\ \hat{m})$, ($\tau^0_{||}\hat{m}\ \times$ $(-\rm\hat{y}$ $\times\ \hat{m})$ for Pt), so that for an in-plane magnetization $\tau_{||}(\phi) = \tau^0_{||} \cos\phi$, while the out-of-plane torque can be the sum of a field-like spin-orbit torque $\tau^0_\text{FL} \hat{m} \times \rm\hat{y}$ and the torque due to the Oersted field $\tau^0_\text{Oe} \hat{m} \times \rm\hat{y}$, both of which give $\tau_{\perp}(\phi) = \tau^0_{\perp} \cos\phi$.
Therefore in this simple case, both $S_\text{XX}$ and $A_\text{XX}$ are proportional to $\sin(2\phi)\cos(\phi)$. For a low-symmetry material like ZrTe$_3$, current-induced spins oriented in the $\hat{z}$ direction are also allowed by symmetry if there is a component of current perpendicular to the mirror plane.\textsuperscript{\cite{MacNeill2017_STFMR, Macneill2017b, Stiehl2019, He2020, Liu2020}} This allows for an in-plane field-like torque of the form $-\tau^{\text{FL}}_{||} \hat{m} \times \rm\hat{z}$ and an out-of-plane anti-damping torque of the form $\tau^{\text{DL}}_{\perp} \hat{m} \times (\rm\hat{z}$ $\times\ \hat{m})$ so that for an in-plane magnetization $\tau_{||}(\phi) = \tau^0_{||} \cos\phi - \tau^{\text{FL}}_{||}$ and $\tau_{\perp}(\phi) = \tau^0_{\perp} \cos\phi + \tau^{\text{DL}}_{\perp}$. Consequently, the amplitudes of the ST-FMR components can have the angular dependence
\begin{equation}
S_{\text{XX}} = \sin(2\phi)(V_\text{S} \cos(\phi) + V_\text{T})
\label{eqn:VphiS}
\end{equation}
\begin{equation}
A_{\text{XX}} = \sin(2\phi)(V_\text{A} \cos(\phi) + V_\text{B})
\label{eqn:VphiA}
\end{equation}
where $V_\text{S}$, $V_\text{T}$, $V_\text{A}$, and $V_\text{B}$ are constants corresponding to the strength of the  $\tau^0_{||}$, $\tau^\text{FL}_{||}$, $\tau^0_{\perp}$ and $\tau^\text{DL}_{||}$ torques respectively.
Fits to the angular dependence for 3 nm and 66 nm thick ZrTe$_3$ devices in \textbf{Figure \ref{AMR-ST-FMR-SA}} show good agreement with the expected angular dependence. For the thinner ZrTe$_3$ layer we observe a non-zero value of $V_T$ corresponding to an unconventional in-plane field-like torque. An unconventional out-of-plane anti-damping torque is allowed by symmetry, but is not evident in the angular fits for either sample.

\vspace{0.5cm}
If one assumes that the out-of-plane field-like spin-orbit torque is negligible relative to the Oersted torque (a good assumption at least for the thicker ZrTe$_3$ layers, see the Supporting Information), one can calculate the efficiencies corresponding to the in-plane torques $\tau^0_{||}$ and $\tau^{\text{FL}}_{||}$ by using the Oersted torque to calibrate the charge current density $J_\text{C}$ in the ZrTe$_3$ layer\textsuperscript{\cite{Liu2011}}
\begin{equation}
    \xi_{||}^{\text{DL}} \equiv \frac{2 \text{e} M_s t_{\text{mag}}}{\upgamma \hbar J_c} \tau^0_{||} = \frac{V_\text{S}}{V_\text{A}}\frac{\text{e}\upmu_0M_st\rm_{ZrTe_3}t_{\text{mag}}}{\hbar}\sqrt{1 + \frac{\upmu_0 M_\text{eff}}{B_0}}
    \label{eqn:xi_0DL}
\end{equation} 
\begin{equation}
    \xi^{\text{FL}}_{||} \equiv \frac{2 \text{e}  M_\text{s} t_{\text{mag}}}{\upgamma \hbar J_c} \tau^{\text{FL}}_{||} = \frac{V_\text{T}}{V_\text{A}}\frac{\text{e}\upmu_0 M_\text{s} t\rm_{ZrTe_3}t_\text{mag}}{\hbar}\sqrt{1 + \frac{\upmu_0 M_\text{eff}}{B_0}}.
    \label{eqn:xi_FLIP}
\end{equation} 
Here $M_\text{s}$ is the saturation magnetization, $t\rm_{ZrTe_3}$ is the ZrTe$_3$ thickness, and $t_\text{mag}$ is the Py thickness. We have also checked this approach for selected sample thicknesses by directly calibrating the microwave current in the sample using a vector network analyzer (see Supporting Information), rather than computing $V_S/V_A$, and the conclusions are the same.

\begin{figure}[htpb]
    \centering
    \subfloat{
    \begin{minipage}{0.5\textwidth}
        \centering
        \includegraphics[width=\textwidth]{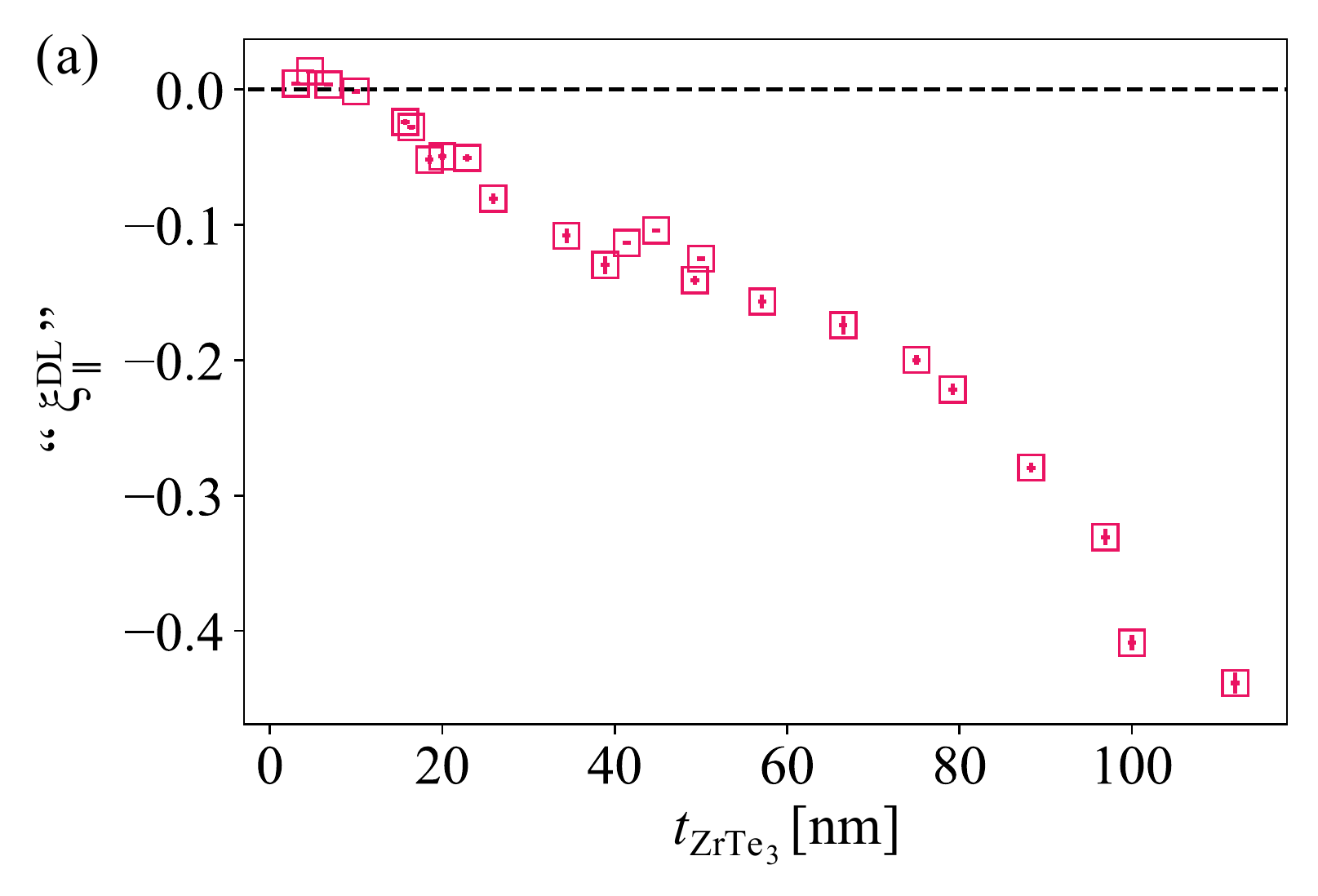}
    \end{minipage}}
    \subfloat{
    \begin{minipage}{0.5\textwidth}
        \centering
        \includegraphics[width=\textwidth]{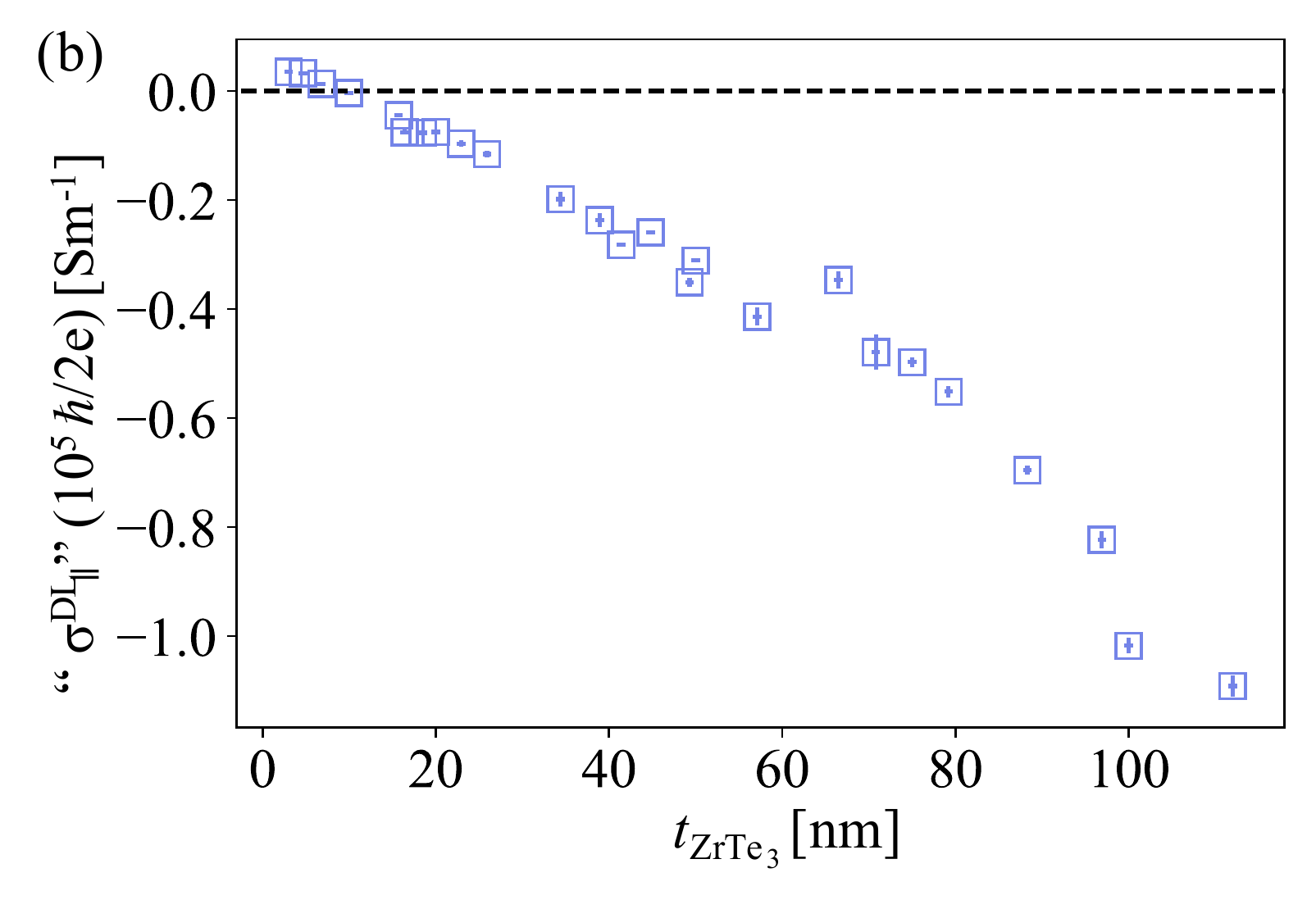}
    \end{minipage}}\\
    \subfloat{
    \begin{minipage}{0.5\textwidth}
        \centering
        \includegraphics[width=\textwidth]{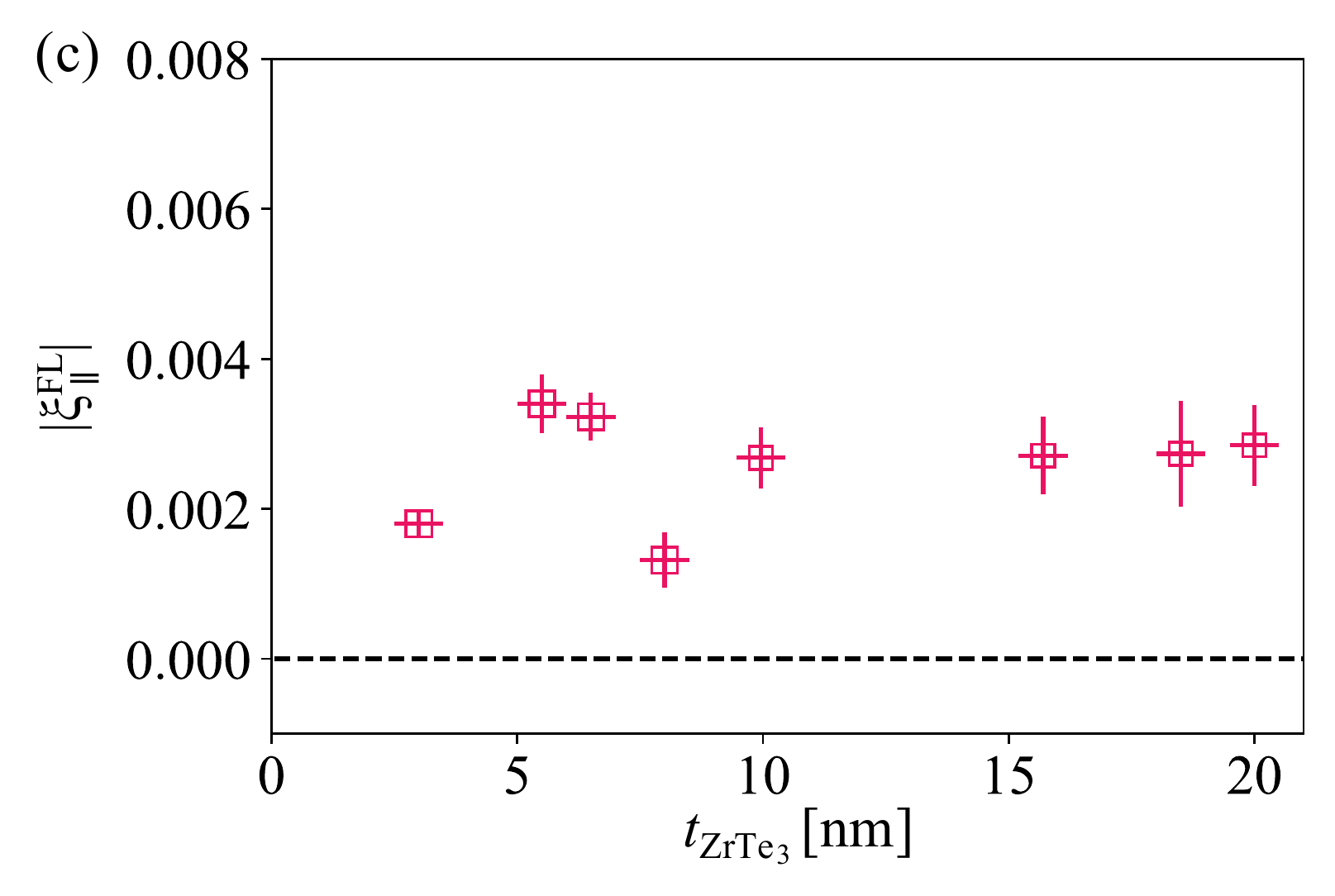}
    \end{minipage}}
    \subfloat{
    \begin{minipage}{0.5\textwidth}
        \centering
        \includegraphics[width=\textwidth]{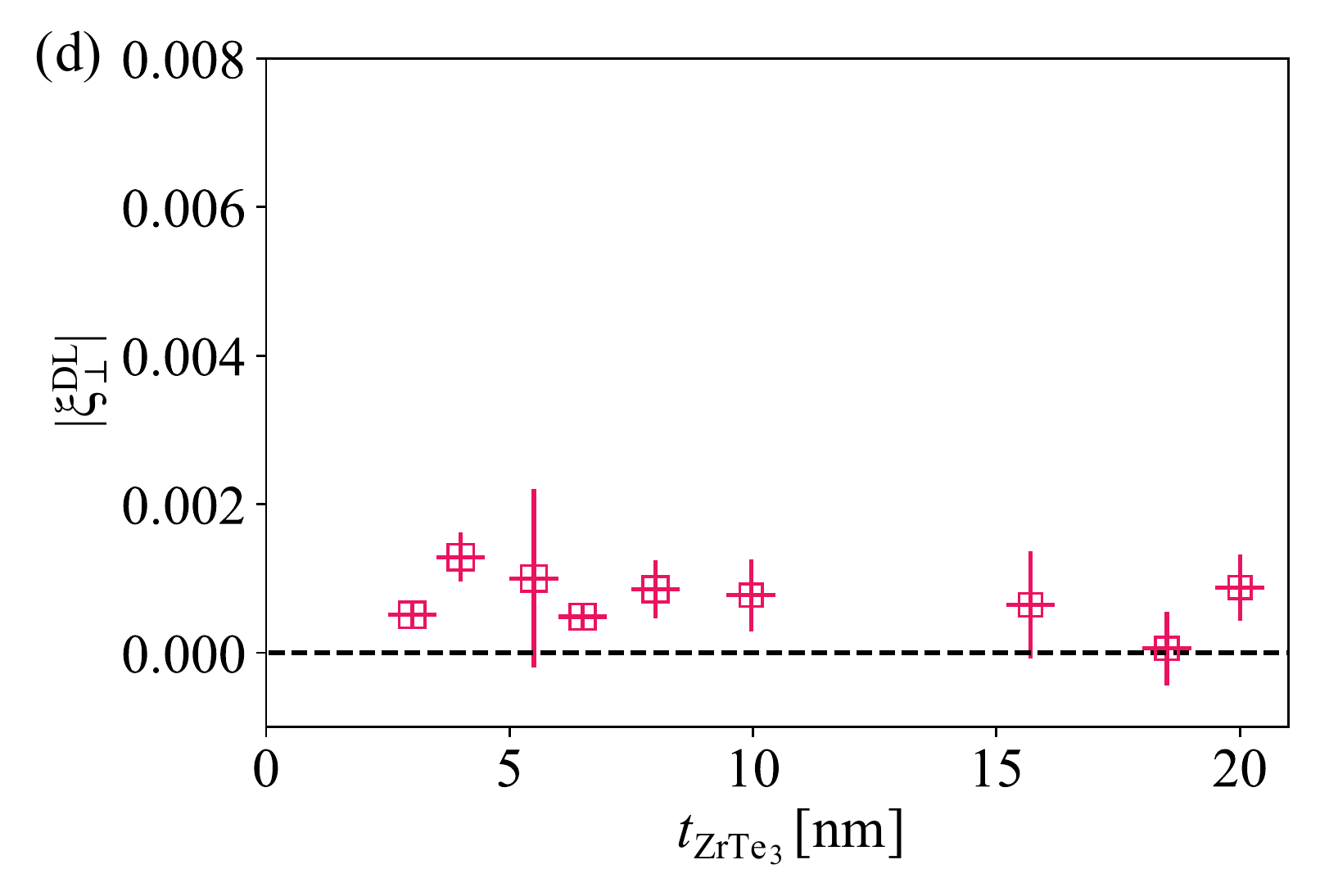}
    \end{minipage}}
    \caption{Thickness dependence of the apparent (a) in-plane anti-damping spin torque efficiency and (b) spin Hall conductivity based on the incorrect standard analysis that neglects artifact from spin pumping and resonant heating. (c) Unconventional in-plane field-like spin torque efficiencies, (d) Unconventional out-of-plane anti-damping spin torque efficiencies.}
    \label{fig:xi}
\end{figure}

\vspace{0.5cm}
The results of this (incorrect) standard analysis, which neglects artifact effects, are shown in \textbf{Figure\ \ref{fig:xi}}a as a function of the thickness of the ZrTe$_3$ layer.  For the thinnest ZrTe$_3$ layers, the standard in-plane anti-damping torque efficiency $\xi_{||}^\text{DL}$ is weakly positive, with a value $\xi_{||}^\text{DL} = 0.015 \pm 0.002$ for the 3 nm ZrTe$_3$ layer, but as a function of increasing ZrTe$_3$ thickness it becomes negative, with strong thickness dependence through 100 nm.  At the largest ZrTe$_3$ thicknesses, the apparent magnitude of $\xi_{||}^\text{DL}$ appears to become extremely large $|\xi_{||}^\text{DL}| >$ 0.4, even larger than the value for pure W.\textsuperscript{\cite{Pai2012}} The corresponding torque conductivity $\sigma_{||}^\text{DL} = \xi_{||}^\text{DL}\sigma_\text{XX}$ (Figure \ref{fig:xi}b), also exhibits the unusual thickness dependence and very large apparent values, on the same order of magnitude as Pt for the thickest ZrTe$_3$ samples, $|\sigma_{||}^\text{DL}| >$ 10$^5$ ($\frac{\hbar}{2e}$) Sm$^{-1}$.\textsuperscript{\cite{Zhu2019}} The unconventional in-plane fieldlike torques have values around $|\xi_{||}^\text{FL}| = 0.003 \pm 0.001$ and remain largely independent of ZrTe$_3$ thickness (Figure \ref{fig:xi}c), while the out-of-plane anti-damping torques are negligible (Figure \ref{fig:xi}d).

\vspace{0.5cm}
This standard analysis is incorrect because of the neglect of the artifacts from spin pumping\textsuperscript{\cite{Tserkovnyak2002, Tserkovnyak2002b, Mosendz2010}} and resonant heating\textsuperscript{\cite{Holanda2017}}.  The Oersted torque generated by the charge current in the ZrTe$_3$ layer is proportional to charge current density in the layer times the ZrTe$_3$ layer thickness, while the spin-orbit torques are proportional only to the charge current density. Consequently, as the ZrTe$_3$ layer thickness increases the Oersted torque increasingly dominates over the spin-orbit torques.  The large Oersted torque generates large precession amplitudes, and hence increased signals due to spin pumping and resonant heating, relative to the mixing voltages generated by the spin-orbit torques.  These artifact voltages produce a symmetric resonant peak shape, and when the ST-FMR resonant voltage is detected only in the longitudinal direction they have the same angular dependence as the signal from the in-plane anti-damping spin-orbit torque, and hence cannot be distinguished from the mixing signal due to $\tau^0_{||}$.  For the sign of $R_{\text{AMR}}$ in Py devices, the sign due to the spin-pumping signal is opposite to the rectification signal due to $\tau^0_{||}$,\textsuperscript{\cite{Karimeddiny2020}} consistent with Figure \ref{fig:xi}a, and the magnitude of the signals we observe in Figure \ref{fig:xi}a are also fully consistent with expectations for the spin pumping + inverse spin Hall effect (see Supporting Information).

\subsection{Using ST-FMR in the Hall geometry to separate the spin-orbit-torque signal from artifacts}
When ST-FMR signals are detected in the Hall geometry in addition to the standard longitudinal geometry, the angular dependence of artifacts due to spin-pumping and resonant heating is no longer identical to the Hall-detected spin-orbit mixing signal, allowing the different signals to be separated.\textsuperscript{\cite{Karimeddiny2020}} We perform the Hall-detected measurements using the sample geometry shown in \textbf{Figure \ref{fig:schematic}}b. We detect both the transverse and longitudinal voltage signals at the same time using two lock-in amplifiers registered to the same kHz-frequency amplitude modulation of the microwave-frequency applied current. Neither the device fabrication nor the measurements themselves therefore take more time than conventional longitudinal ST-FMR measurements. To obtain quantitative values of the current-induced torques, we calibrate the microwave current $\Irf$ for each device using a vector network analyzer (see Supporting Information).

\begin{figure}[htpb]
    \centering
    \subfloat{
    \begin{minipage}{0.5\textwidth}
        \centering
        \includegraphics[width=\textwidth]{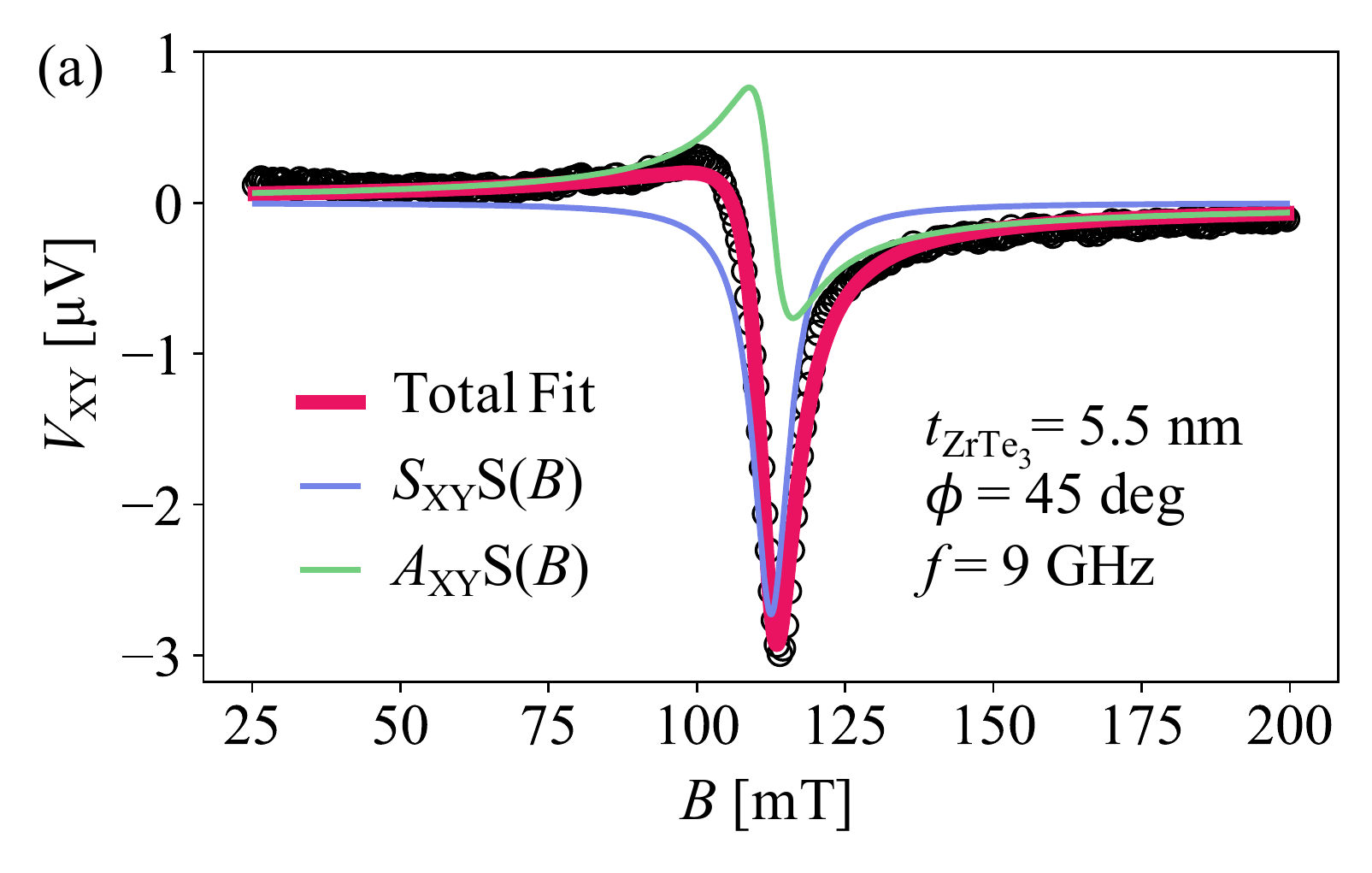}
    \end{minipage}\hfill}
    \subfloat{
    \begin{minipage}{0.5\textwidth}
        \centering
        \includegraphics[width=\textwidth]{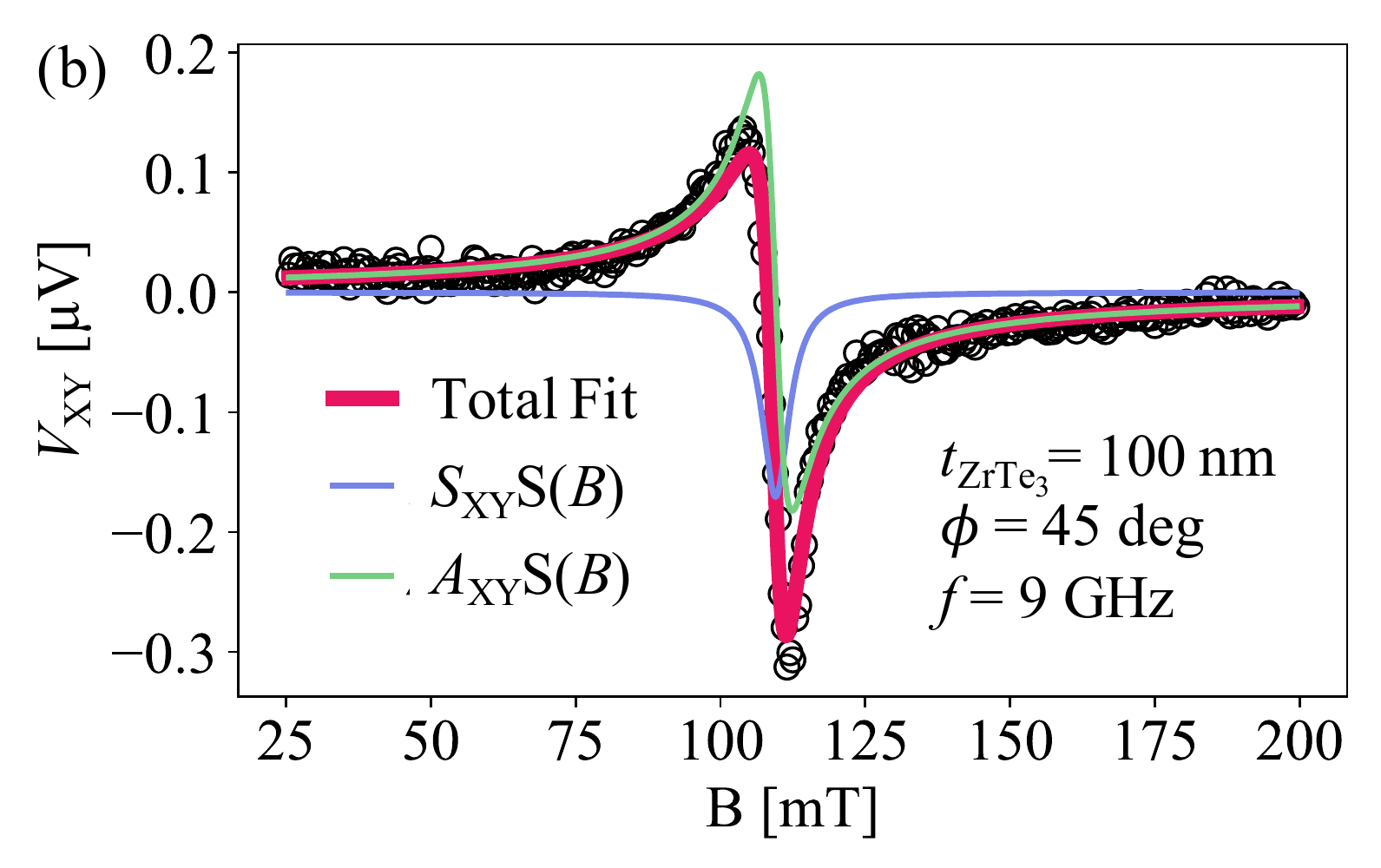}
    \end{minipage}}
    \caption{Hall ST-FMR signal as a function of applied magnetic field at a fixed angle of 45$^{\circ}$ for a 5.5 nm and a 100 nm thick ZrTe$_3$ flake, along with fits to the sum of antisymmetric and symmetric Lorentzian line shapes. }
    \label{AMR-ST-FMR-res}
\end{figure}

\vspace{0.5cm}
We follow the analysis procedure described in reference \cite{Karimeddiny2020} with the addition of contributions from the unconventional torques $\tau^\text{FL}_{||}$ and $\tau^\text{DL}_{\perp}$. The transverse ST-FMR signal can again be separated into symmetric and anti-symmetric Lorentzian components, $V_\text{XY}$ = $S_\text{XY}$S($B$) + $A_\text{XY}$A($B$) (\textbf{Figure\ \ref{AMR-ST-FMR-res}}). 

\vspace{0.5cm}
The artifact-induced voltages have the angular dependence:
\begin{align}
    \begin{split}
        \label{eq:SnAwithAngles}
            V_\text{art} &= E^{0}_\text{art} \text{cos}^2 \phi
            \begin{cases}
            L\ \text{sin}\ \phi & \text{longitudinal} \\
            W\ \text{cos}\ \phi & \text{transverse}
            \end{cases}\\
        &= \frac{E^{0}_\text{art}}{2}
        \begin{cases}
        L\ \text{sin}\ 2\phi\ \text{cos}\ \phi & \text{longitudinal} \\
        W\ (\text{cos}\ 2\phi \text{cos}\ \phi + \text{cos}\ \phi) & \text{transverse}
        \end{cases}
\end{split}
\end{align} where $L$ is the length of the device (in the longitudinal direction) and $W$ is the width, and we have used the relations cos$^2$$\phi$sin$\phi$ = (sin2$\phi$cos$\phi$)/2 and cos$^3$$\phi$ = (cos$\phi$ + cos2$\phi$cos$\phi$)/2. We add these contributions from the artifact-induced voltages $V_\text{art}$ to the mixing voltages from current-induced magnetic precession. The mixing voltages in the Hall geometry have contributions from modulation of both the planar and anomalous Hall effects, $R_\text{XY}$($\phi$) = $R_\text{PHE}$ sin$^2(\theta)$sin($\phi$)cos($\phi$) + R$_\text{AHE}$ cos($\theta$), times angular dependence associated with $\vec{\tau}_{||}(\phi)$ and  $\vec{\tau}_{\perp}(\phi)$ (see the derivation in \cite{Karimeddiny2020}). Here $\theta$ is the tilt angle of the magnetization relative to the out-of-plane direction.  The general form of the $\phi$ dependence for the case in which current-generated spins are allowed in both the $\hat{y}$ and $\hat{z}$ directions is
\begin{align}
\begin{split}\label{eq:SnAwithAngles}
        S_\text{XX}(\phi) &=  S_\text{XX}^\text{AMR/art}\sin 2\phi\cos\phi + S_\text{XX}^{||,\text{FL}} \sin 2\phi, \\
		A_\text{XX}(\phi) &= A_\text{XX}^\text{AMR} \sin 2\phi\cos\phi + A_\text{XX}^{\perp,\text{DL}} \sin 2\phi,\\
		S_\text{XY}(\phi) &= S_\text{XY}^\text{PHE/art}  \cos2\phi\cos\phi + S_\text{XY}^\text{AHE/art}\cos\phi + S_\text{XY}^{||,\text{FL,PHE}}\cos2\phi +S_\text{XY}^{\perp,\text{DL,AHE}}, \\
		A_\text{XY}(\phi) &= A_\text{XY}^\text{PHE}  \cos2\phi\cos\phi
		+ A_\text{XY}^\text{AHE} \cos\phi + A_\text{XY}^{\perp,\text{DL,PHE}}\cos2\phi + A_\text{XY}^{||,\text{FL,AHE}}
\end{split}
\end{align}
with the amplitude coefficients
\begin{align}
    \begin{split}
        \label{AScoefs}
        S_\text{XX}^\text{AMR/art} &= \frac{\Irf}{2\alpha\omega^+} R_\text{AMR}\tau^0_{||} - \frac{L}{2}  E^0_\text{art}  \\
                            &\equiv S_\text{XX}^\text{AMR} + V_\text{art}\\
        S_\text{XX}^{||,\text{FL}} &= -\frac{\Irf}{2\alpha\omega^+} R_\text{AMR}\tau^{\text{FL}}_{||}  \\
		A_\text{XX}^{\text{AMR}} &= \frac{\Irf}{2\alpha\omega^+} R_\text{AMR}\frac{\omega_2}{\omega}\tau^0_{\perp}\\
		A_\text{XX}^{\perp,\text{DL}} &= \frac{\Irf}{2\alpha\omega^+} R_\text{AMR}\frac{\omega_2}{\omega}\tau^{\text{DL}}_{\perp}  \\
		S_\text{XY}^{\text{PHE/art}} &= -\frac{\Irf}{2\alpha\omega^+} R_\text{PHE}\tau^0_{||} - \frac{W}{2} E^0_\text{art} \\
		A_\text{XY}^{\text{PHE}} &= -\frac{\Irf}{2\alpha\omega^+}R_\text{PHE}\frac{\omega_2}{\omega}\tau^0_{\perp}\\
		S_\text{XY}^{\text{AHE/art}} &=  \frac{\Irf}{2\alpha\omega^+}R_\text{AHE}\tau^0_{\perp} - \frac{W}{2} E^0_\text{art} \\
		A_\text{XY}^{\text{AHE}} &= -\frac{\Irf}{2\alpha\omega^+} R_\text{AHE}\frac{\omega_1}{\omega} \tau^0_{||} \\
		S_\text{XY}^{||,\text{FL,PHE}} &= \frac{\Irf}{2\alpha\omega^+} R_\text{PHE}\tau^{\text{FL}}_{||}\\
		A_\text{XY}^{\perp,\text{DL,PHE}} &= -\frac{\Irf}{2\alpha\omega^+}R_\text{PHE}\frac{\omega_2}{\omega}\tau^{\text{DL}}_{\perp} \\
		S_\text{XY}^{\perp,\text{DL,AHE}} &=  \frac{\Irf}{2\alpha\omega^+}R_\text{AHE}\tau^{\text{FL}}_{\perp} \\
		A_\text{XY}^{||,\text{FL,AHE}} & = \frac{\Irf}{2\alpha\omega^+} R_\text{AHE}\frac{\omega_1}{\omega} \tau^{\text{FL}}_{||}.
    \end{split}
\end{align}
Here we have assumed that the unconventional components of the spin Hall conductivity are sufficiently small that they do not influence the angular dependence of the artifact voltage.

\begin{figure}[htpb]
        \centering
        \includegraphics[width=\textwidth]{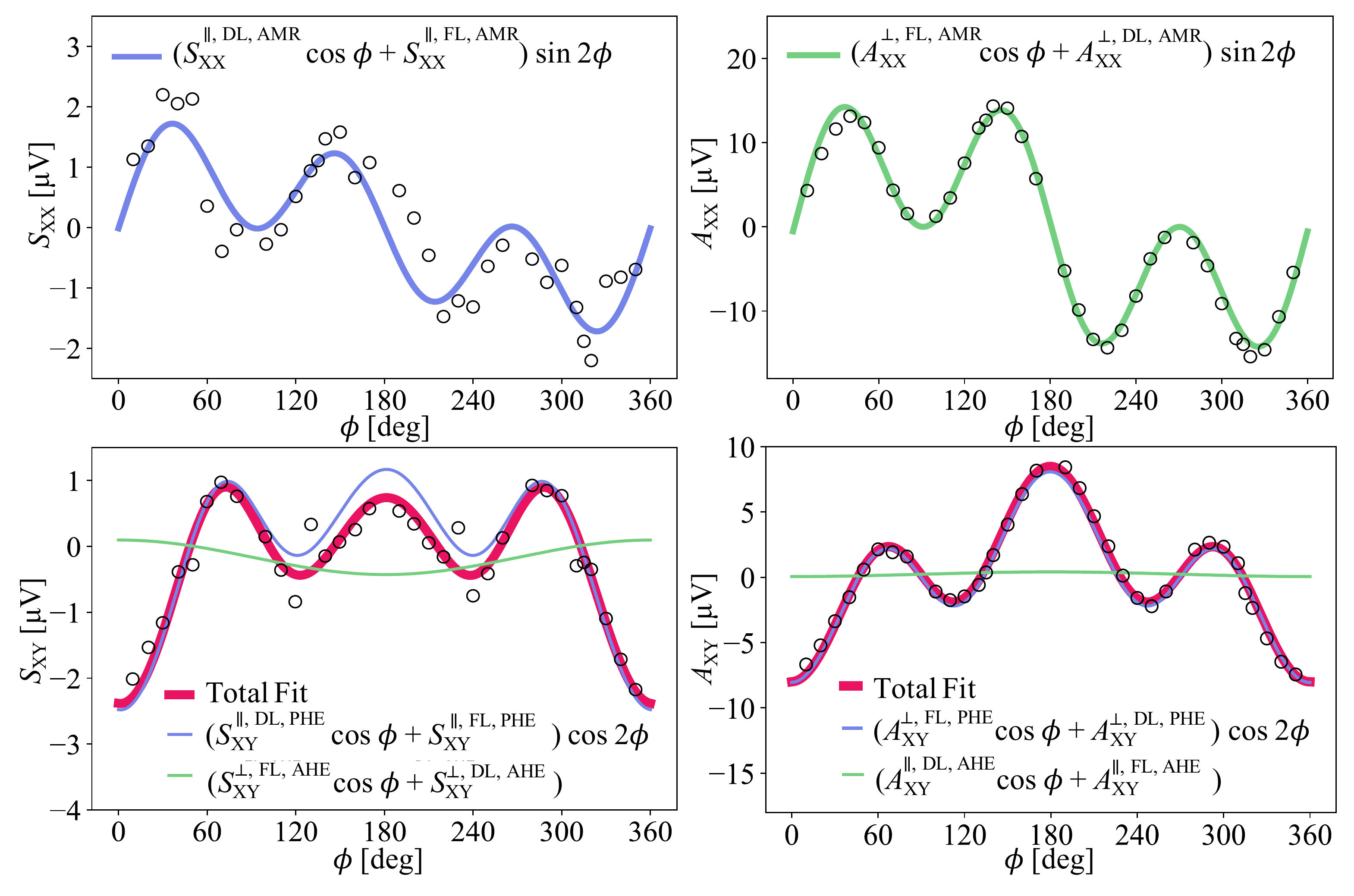}
    \caption{Angular dependences of symmetric and antisymmetric resonance components of longitudinal ($S_\text{XX}$, $A_\text{XX}$) and Hall ($S_\text{XY}$, $A_\text{XY}$) mixing voltages for a 5.5 nm thick ZrTe$_3$ device. }
    \label{Hall-ST-FMR-SA}
\end{figure}

\begin{figure}[htpb]
    \centering
    \subfloat{
    \begin{minipage}{0.5\textwidth}
        \centering
        \includegraphics[width=\textwidth]{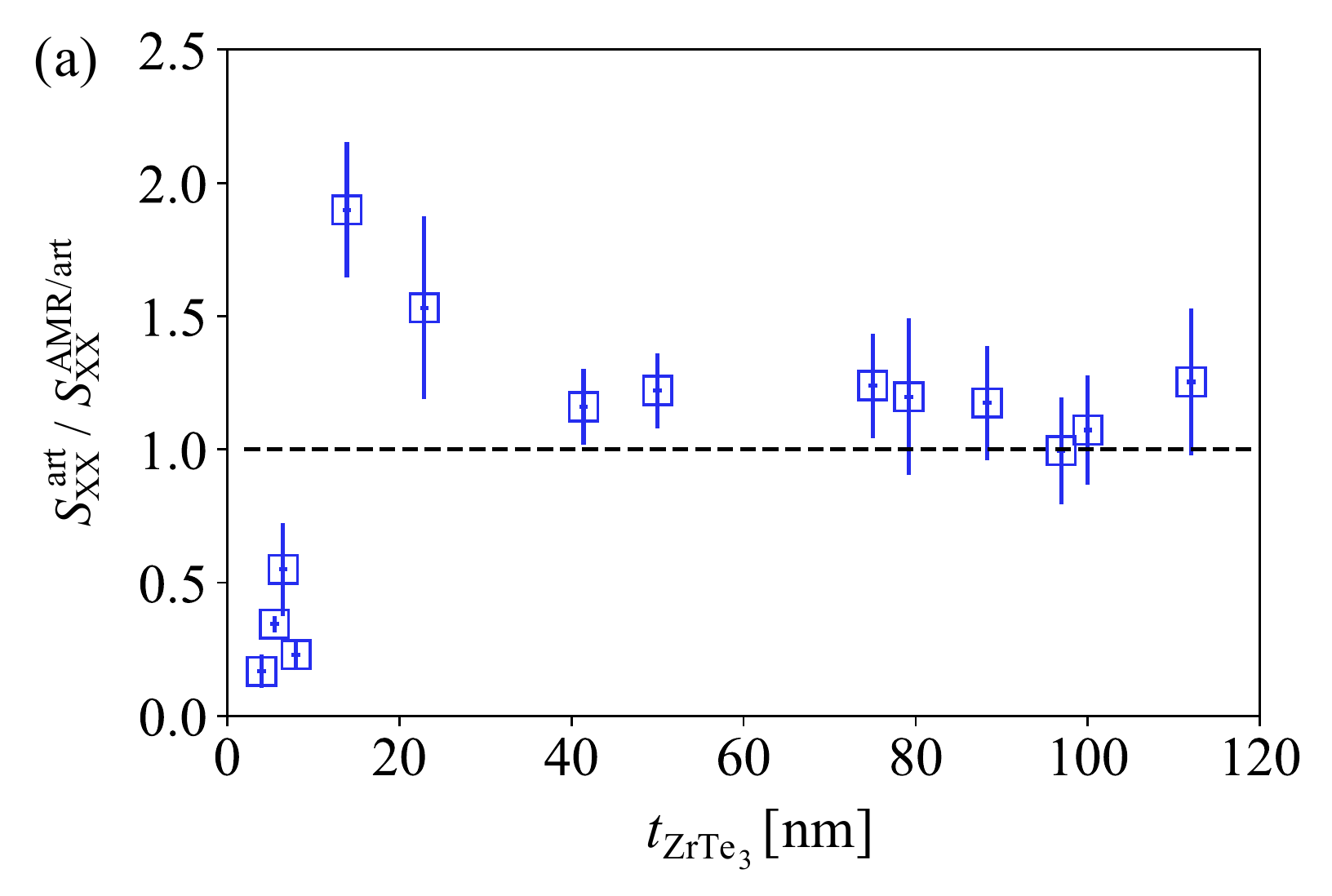}
    \end{minipage}\hfill}
    \subfloat{
    \begin{minipage}{0.5\textwidth}
        \centering
        \includegraphics[width=0.95\textwidth]{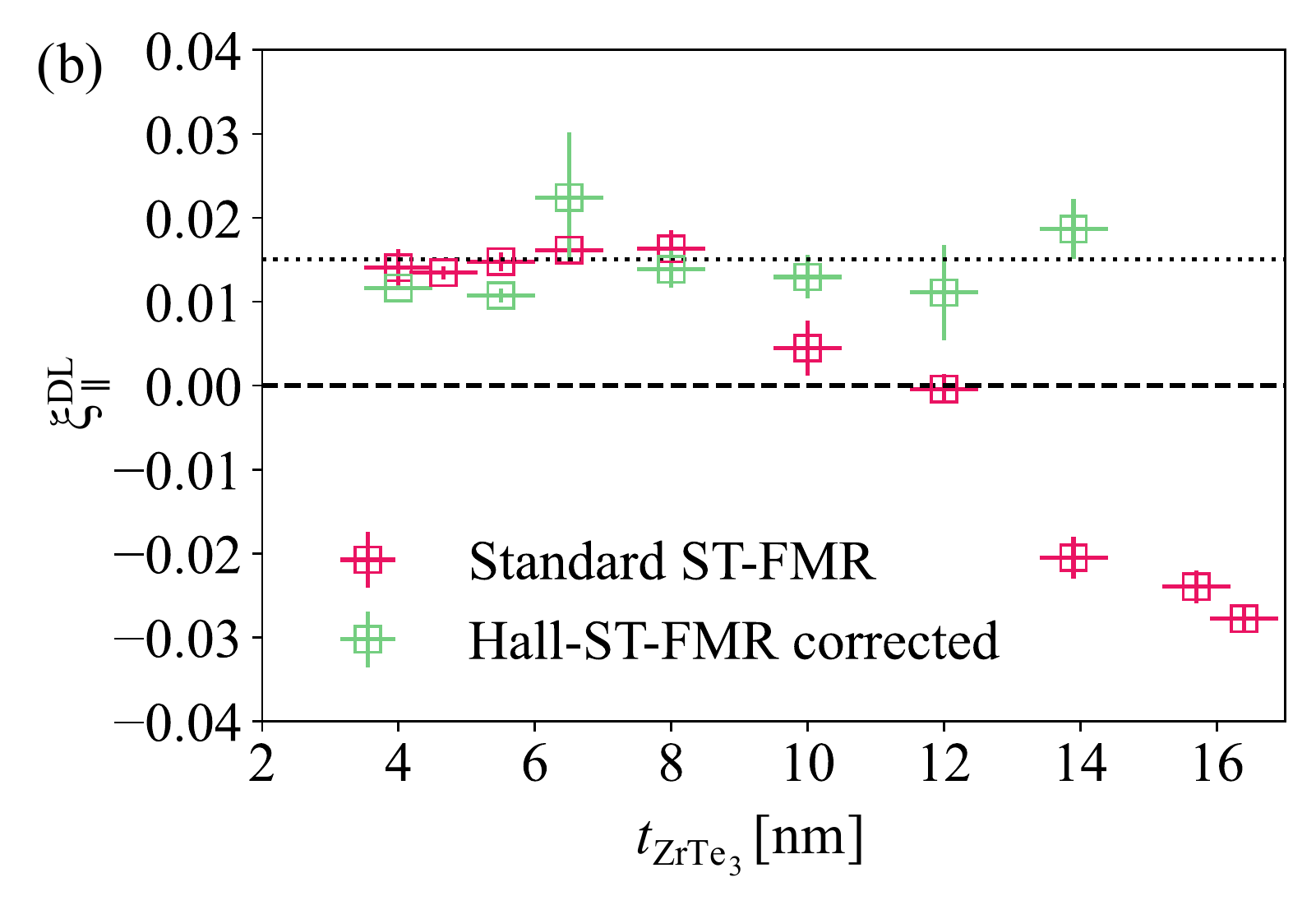}
    \end{minipage}}\\
    
    \subfloat{
    \begin{minipage}{0.5\textwidth}
        \centering
        \includegraphics[width=\textwidth]{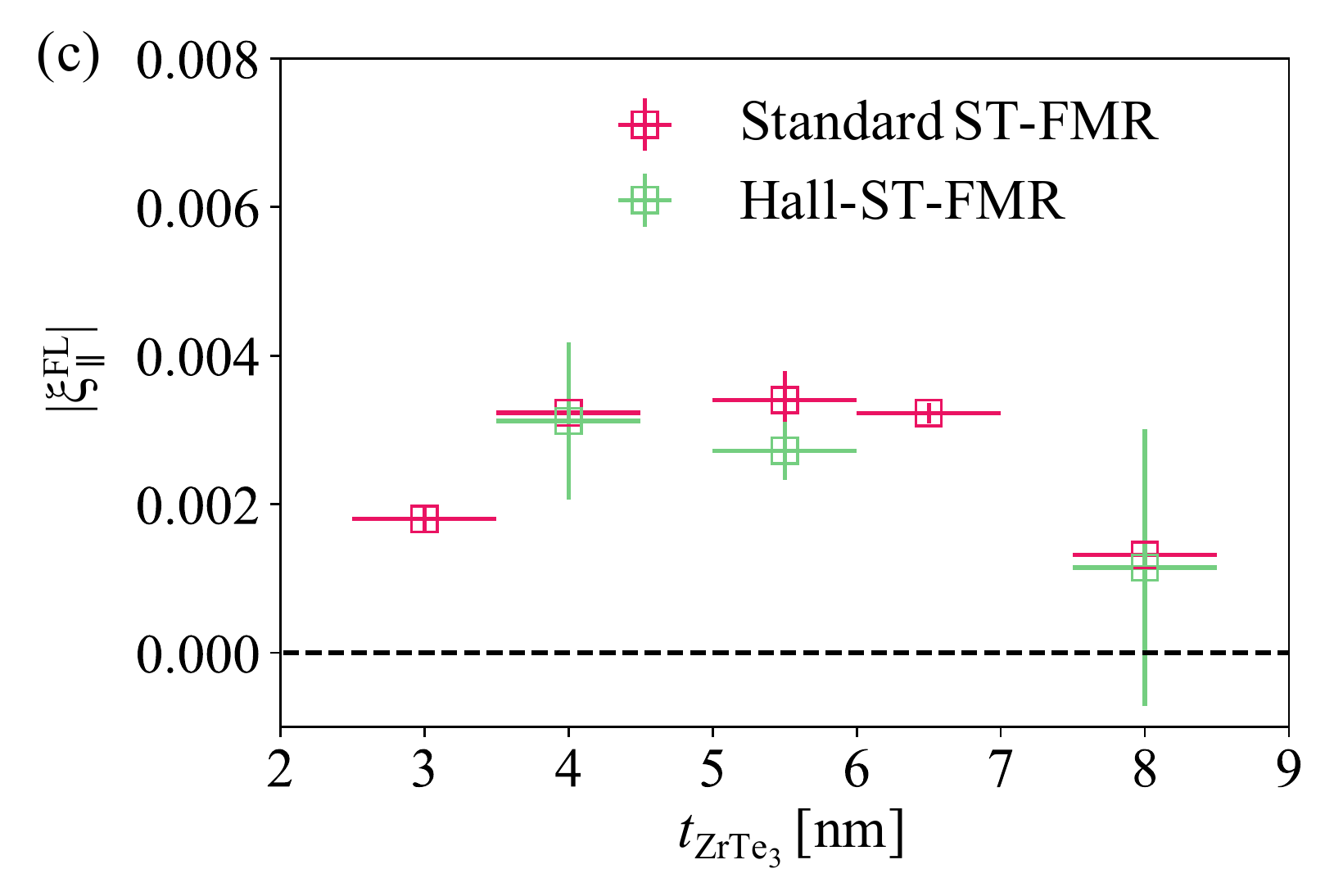}
    \end{minipage}\hfill}
    \subfloat{
    \begin{minipage}{0.5\textwidth}
        \centering
        \includegraphics[width=0.98\textwidth]{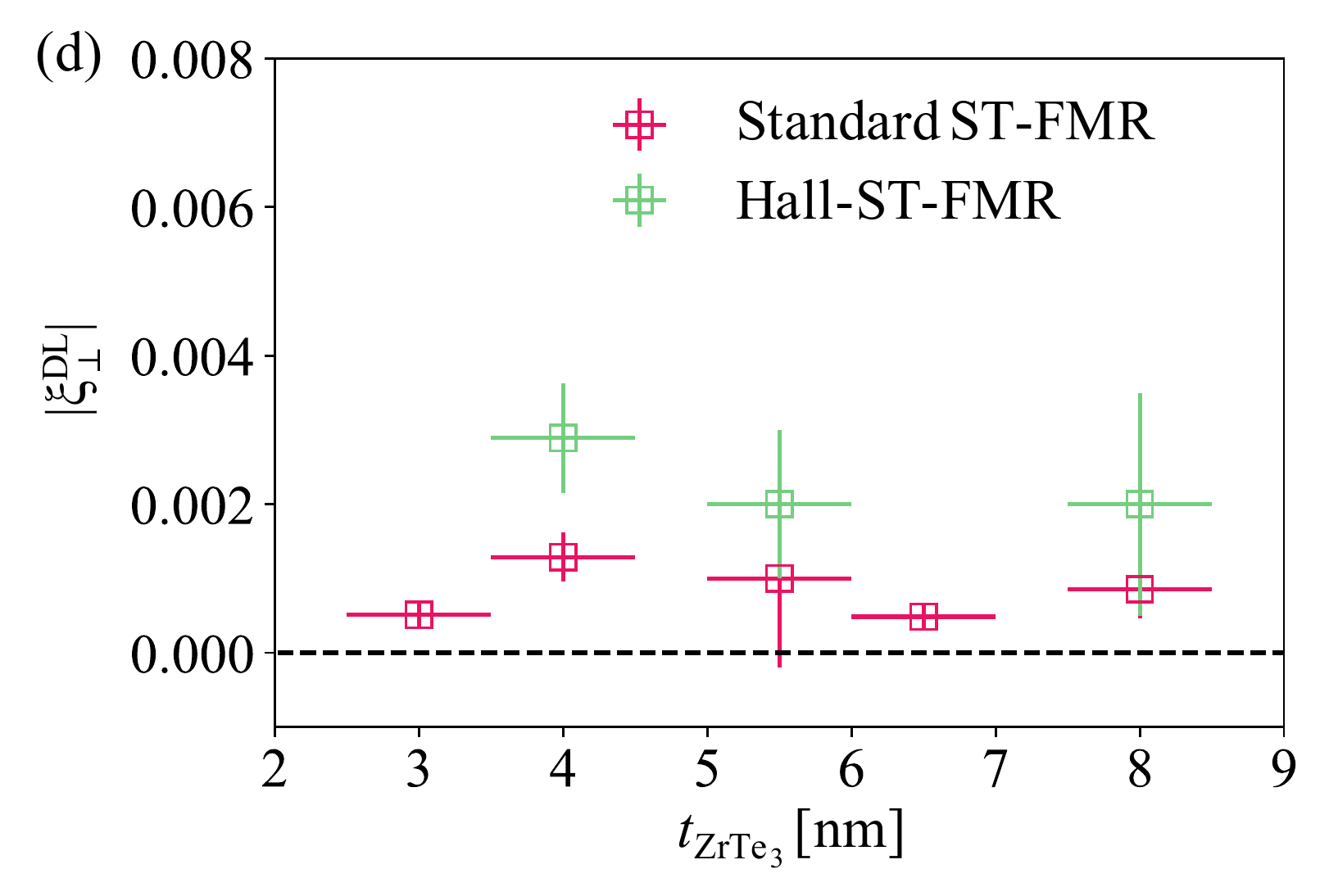}
    \end{minipage}}
    \caption {(a) Ratio of the artifact voltage $S^{\text{art}}_\text{XX}$ to the measured symmetric longitudinal ST-FMR signal amplitude $S^{\text{AMR/art}}_{\text{XX}}$, showing that the artifact dominates the signal for for $t\rm_{ZrTe_3}$ $>$ 10 nm. (b)  Artifact-corrected anti-damping torque efficiency from Hall ST-FMR, showing negligible thickness dependence and absence of a sign change as $t\rm_{ZrTe_3}$ increases. Spin torque efficiencies from the Hall ST-FMR measurements for (c) the unconventional in-plane field-like torque and (d) the unconventional out-of-plane anti-damping torque, showing small values in agreement with the standard ST-FMR measurements for these torque components.}
    \label{fig:xi_corr}
\end{figure}

\vspace{0.5cm}
\textbf{Figure \ref{Hall-ST-FMR-SA}} shows representative fits of Equation\ (\ref{eq:SnAwithAngles}) to the measured angular dependence of $S_\text{XX}$, $A_\text{XX}$, $S_\text{XY}$, and $A_\text{XY}$ for a ZrTe$_3$ (5 nm)/Py (6 nm) sample.  We find good agreement.  We can determine the artifact electric field by first calculating the ratio $\eta \equiv (\tau^0_{||}/\tau^0_\perp)\sqrt{\omega_1/\omega_2}$ in two different ways by employing the pair of parameters $S$ and $A$ associated with each of the AMR, PHE, and AHE corresponding to $\tau^0_{||}$ and $\tau^0_\perp$
\begin{equation}
\eta = \frac{-\Aa}{\Sa + W (E_\text{art}/2)} = \frac{\Sp+ W (E_\text{art}/2)}{\Ap}
\label{AHE}
\end{equation}
\begin{equation}
\eta = \frac{-\Aa}{\Sa + W (E_\text{art}/2)} = \frac{\Sm + L  (E_\text{art}/2)}{\Am}
\label{PHE}
\end{equation}
Using the measured amplitude coefficients, one can solve for $E_\text{art}$ using either Equation\ (\ref{AHE}) or (\ref{PHE}) and check consistency (see Supporting Information for details about the sign in the quadratic formula). We find that these values do agree to within experimental uncertainty. \textbf{Figure \ref{fig:xi_corr}}a shows the ratio of the artifact voltage $S^{\text{art}}_\text{XX} = L (E_\text{art}/2)$ to the total measured value of $S^{\text{AMR/art}}_\text{XX}$.  We see for ZrTe$_3$ layers thicker than 10 nm that this ratio is close to 1, meaning that the longitudinal symmetric ST-FMR signals are completely dominated by the artifact voltage.

\vspace{0.5cm}
Figure \ref{fig:xi_corr}b shows the results of the Hall ST-FMR analysis for the efficiency of the conventional in-plane anti-damping spin-orbit torque, separated from the artifact signals. To arrive at these values, we determine $\tau^0_{||}$ by subtracting the artifact voltage from the measured values of $S_{XX}^\text{AMR/art}$ in Equation (\ref{AScoefs}), measuring $\alpha$ from the ST-FMR linewidths, $M_\text{s}$ from the frequency dependence of $B_0$ (assuming $M_\text{s} \approx M_\text{eff}$), and calibrating $\Irf$ for each device using a vector network analyzer. The in-plane anti-damping spin-torque efficiency $\xi\rm_{||}^{DL}$ is then determined from $\tau^0_{||}$ using Equation\ (\ref{eqn:xi_0DL}) using a parallel-resistor model to estimate the charge current density within the ZrTe$_3$ layer.  

\vspace{0.5cm}
We obtain the efficiency $\xi\rm_{||}^{DL}$ = 0.014 $\pm$ 0.004 as shown by the dotted line in figure\ \ref{fig:xi_corr}, largely independent of device thickness for $t\rm_{ZrTe_3}$ $<$ 15 nm. Beyond ZrTe$_3$ thicknesses of 15 nm, the artifact voltages are too large to make an accurate determination of the spin-orbit torque, but it is clear that the apparent thickness dependence of this efficiency in the range $t\rm_{ZrTe_3}$ $>$ 15 nm in Figure 5a is due entirely to the effects of the artifact voltages. The unconventional torques from the Hall geometry (Figure \ref{fig:xi_corr}c,d), also remain largely independent of thickness, with values that are little-changed from the conventional ST-FMR analysis.

\section{Conclusion}
We have used ST-FMR to investigate the spin-orbit torques generated by exfoliated flakes of the low-symmetry semi-metal ZrTe$_3$ for a wide range of layer thicknesses in ZrTe$_3$/Py(6 nm) devices. We find that the ``standard'' ST-FMR analysis, which neglects the effects of artifacts due to spin pumping and resonant heating, gives incorrect values for the in-plane anti-damping torque efficiency $\xi_{||}^\text{DL}$ of ZrTe$_3$ layers thicker than about 10 nm.  For the thickest layers, this incorrect standard analysis can overestimate the magnitude of $\xi_{||}^\text{DL}$ by as much as a factor of 30, and it indicates an unphysical strong dependence of the torque efficiency on layer thickness.  ST-FMR measurements in the Hall geometry demonstrate that this strong apparent thickness dependence is due entirely to artifacts from spin pumping and/or resonant heating, not a true dependence of the spin-orbit torque on layer thickness. For ZrTe$_3$, the Hall ST-FMR measurements yield torque efficiencies  $\xi_{||}^\text{DL}$ = 0.014 $\pm$ 0.004 for the conventional in-plane anti-damping torque and $|\xi_{||}^{\text{FL}}|$ = 0.003 $\pm$ 0.001 for the unconventional in-plane field-like torque.  The unconventional spin-orbit torques in ZrTe$_3$ are similar to strained NbSe$_2$ in that the in-plane field like torque is non-zero while the out-of-plane anti-damping torque is negligible, while low-symmetry WTe$_2$ and MoTe$_2$ are different in that the out-of-plane anti-damping torque is non-zero.

\vspace{0.5cm}
We can make an estimate for when artifact voltages cannot be neglected for an arbitrary nonmagnetic (NM) spin-source material by calculating the ratio of the longitudinal spin-pumping voltage $V_\text{sp}$ (Eq.\ (1) in the Supporting Information) divided by the mixing voltage associated with the conventional in-plane antidamping torque $S_\text{XX}$ (Eq.\ (\ref{eqn:Vxxa}) in the main text.)  To simplify the approximation, we consider the case where the Oersted field is the primary driver of the precession that generates the spin-pumping voltage, that the applied magnetic field $B_0$ can be neglected relative to $\mu_0 M_\text{eff}$, and that the scale of the magnetoresistance varies with the NM layer thickness as $R_\text{AMR} \approx G^0_\text{AMR}R^2_\text{tot}$, where $G^0_\text{AMR}$ is the magnetoconductance in the limit of zero NM-metal thickness and $R_\text{tot}$ is the total device resistance. We also assume that the thickness of the NM layer is greater than the spin diffusion length, $t_\text{NM} \gg \lambda_\text{sd}$.  After some algebra, we obtain
\begin{align}
        \label{eqn:VspoverVs}
            \frac{|V_\text{SP}|}{|S_\text{XX}|} &\approx
            \frac{t^2_\text{NM} \lambda_\text{sd} g_\text{eff}^{\uparrow \downarrow}}{\rho_\text{NM} T_\text{int}} \frac{e^2}{4\pi \hbar} 
            \frac{\gamma \mu_0^2 M_\text{s} t_\text{mag}}{\alpha G^0_\text{AMR}}.
\end{align} 
Here $\rho_\text{NM}$ is the resistivity of the nonmagnetic material, $g_\text{eff}^{\uparrow \downarrow}$ is the effective spin-mixing conductance of the interface, and $T_\text{int}$ is an interface transparency for spin currents going from the nonmagnet to the magnet that we approximate $\approx$ 1.  We note that $|V_\text{SP}|/|S_\text{XX}|$ has no dependence on the value of the spin Hall torque efficiency $\xi^\text{DL}_{||}$ as long as this is non-zero.  The relative importance of the spin-pumping artifact grows with layer thicknesses  $\propto t^2_\text{NM}$ and $\propto t_\text{mag}$.  For parameters appropriate for our ZrTe$_3$/Py samples, $\rho_\text{NM}$ = 5.7 x 10$^{-6}$ $\Omega$m, $\mu_0 M_\text{s} = 0.95$ T, $t_\text{mag}$ = 6 nm, $\alpha = 0.01$, and $G^0_\text{AMR}$ = 1 x 10$^{-5}$ $\Omega^{-1}$, and for typical values $g_\text{eff}^{\uparrow \downarrow}$ = 2 x 10$^{19}$ m$^{-2}$ and $\lambda_\text{sd}$ = 2 nm, we obtain  $|V_\text{SP}|/|S_\text{XX}|$ $\approx$ 0.2 for a 10 nm ZrTe$_3$ layer and $|V_\text{SP}|/|S_\text{XX}|$ $\approx$ 20 for a 100 nm device, reasonably consistent with our measurements. If the parameters of the magnetic layer are fixed, then by Eq.\ (\ref{eqn:VspoverVs}) the crossover thickness of the normal metal beyond which spin pumping cannot be neglected scales as 
\begin{align}
    t^\text{crossover}_\text{NM} \propto \sqrt{\frac{\rho_\text{NM} T_\text{int}}{\lambda_\text{sd} g_\text{eff}^{\uparrow \downarrow}}}.
\end{align}
Because of the square root, the crossover thickness is only weakly dependent on the properties of the nonmagnetic layer. We conclude, as a rule of thumb, that the effects of spin-pumping signals should not be ignored in any conventional ST-FMR experiment that employs nonmagnetic layers approaching 10 nm or above.

\vspace{0.5cm}
Several previous experiments studying spin-orbit torques generated by van der Waals layers have performed a conventional ST-FMR analysis, without accounting for the possibility of artifacts due to spin pumping or resonant heating, for devices with layers considerably thicker than 10 nm.\textsuperscript{\cite{Shi2019, Liang2020, Xu2020}} They have reportedly unexpectedly large values for the spin-torque efficiencies compared to thinner layers\textsuperscript{\cite{MacNeill2017_STFMR, Macneill2017b, Stiehl2019}} and strong thickness dependence in the torque efficiency in this large-thickness range beyond 10 nm -- qualitatively similar to what we find in our ZrTe$_3$ samples when we neglect artifacts.  We suggest that these anomalous results are due to spin pumping and/or resonant-heating artifacts, and that these measurements therefore do not provide accurate values of spin-orbit torques. We also suggest that ST-FMR measurements in the Hall geometry should be adopted as a standard technique to allow a clear separation of true spin-orbit-torque signals from these artifacts.

\section{Experimental Section}
\subsection{Device Fabrication and Characterization}
We make spin-orbit-torque test structures by first using the scotch-tape exfoliation method to transfer flakes from bulk ZrTe$_3$ crystals bought commercially from HQ-Graphene, onto high-resistivity silicon/silicon dioxide (300 nm) wafers in a nitrogen glove box with H$_2$O and O$_2$ levels $<$0.5 ppm. For the final stage of exfoliation, scotch-tape covered crystals are transferred into the load lock of a magnetron sputtering system, where pristine surfaces of ZrTe$_3$ are exfoliated under a vacuum of $<$ 10$^{-6}$ torr. We then use grazing angle sputtering to deposit 6 nm of Permalloy (Py, Ni$_{80}$Fe$_{20}$) and 2 nm of aluminum to cap the film.  The aluminum cap is oxidized upon exposure to air. Flakes of appropriate thicknesses are then screened using optical contrast, and selected based on the results of atomic force microscopy (AFM). Figure \ref{fig:ZrTe3Xtal}b shows a schematic of the Py-covered ZrTe$_3$ heterostructure. 

\vspace{0.5cm}
Regions of flakes to be incorporated into devices for study are chosen so that they are smooth ($<$ 0.3 nm roughness) with no mono-layer steps and are free of residue from the exfoliation process. We pattern bars along length of the ZrTe$_3$ nano-ribbons, parallel to the b axis and perpendicular to the mirror plane, with typical dimensions of 6 $\times$ 4 $\upmu$m$^2$, using electron-beam lithography and ion-milling. Electrical contacts for ST-FMR measurements are made using a second round of e-beam lithography and magnetron sputtering of 5 nm Ti/60 nm Pt.  We have studied two types of device geometries: standard ST-FMR bars in which only longitudinal electrical signals can be measured (Figure\ \ref{fig:schematic}(a)) and ST-FMR devices which also contain Hall contacts (with 1 $\upmu$m side wires) (Figure\ \ref{fig:schematic}(b)). The crystallographic orientations of the ZrTe$_3$ in the completed devices are further confirmed using polarized Raman spectroscopy on a WITec Alpha300R confocal Raman microscope fitted with a Thorlabs rotation stage (see Supporting Information).

\subsection{ST-FMR Measurement}
ST-FMR measurements were done using an Agilent E8257C 40 GHz rf power source and Signal Recovery 7265 Lock-in amplifiers for readout of the mixing voltages. RF signals with frequencies 9-12 GHz were input with a maximum power of 10 dBm along the low symmetry b axis of the ZrTe$_3$/Py heterostructures. In-plane magnetic fields were applied on a probe station using a GMW 5201 projected field magnet mounted on x, y and phi motion stages controlled by a Newport ESP300 motion controller. Fields were swept from 0 - 0.27 T at varying in-plane angles ($\phi$) with respect to the current direction while keeping the rf frequency constant. The rf current calibration was done through S$_\text{11}$ and S$_\text{21}$ measurements using an Agilent 8722ES 40 GHz Network Analyzer.

\medskip
\textbf{Supporting Information} \par
Supporting Information is available from the Wiley Online Library or from the author.

\medskip
\textbf{Acknowledgements} \par
We acknowledge helpful discussions with Arnab Bose and Rakshit Jain, and nano-fabrication advice and measurement support from Jeremy Clark and Steve Kriske. Primary support for research expenses came from the US Dept. of Energy (DE-SC0017671). T.M.C.\ led the sample fabrication, measurement, and analysis, supported by the Singapore Agency for Science, Technology, and Research. S.K.\ assisted with calibrations and data analysis, supported by the NSF (DMR-1708499).  Assistance with deposition of the heterostructures was provided by V.G.\ (funded by the AFOSR-MURI project 2DMagic, FA9550-19-1-0390) and J.A.M.\ (funded by Task 2776.047 of ASCENT, one of six centers in JUMP, a Semiconductor Research Corporation program sponsored by DARPA).  The devices were fabricated using the shared facilities of the Cornell NanoScale Facility, a member of the National Nanotechnology Coordinated Infrastructure (supported by the National Science Foundation (NSF), NNCI-1542081) and the facilities of Cornell Center for Materials Research (supported by the NSF, DMR-1719875).

\medskip
\bibliographystyle{MSP}
\bibliography{main}

\newpage
\section{Figures}

\newpage



\section*{Supplementary material}

# Supporting Information for:

# Separation of Artifacts from Spin-Torque Ferromagnetic Resonance Measurements of Spin-Orbit Torque for the Low-Symmetry van der Waals Semi-Metal ZrTe$_3$

*Thow Min Cham,*[1] *Saba Karimeddiny,*[1] *Vishakha Gupta,*[1] *Joseph A. Mittelstaedt,*[1] *and Daniel C.Ralph*[1,2]

[1]Cornell University, Ithaca, NY 14850, USA
[2]Kavli Institute at Cornell, Ithaca, NY 14853, USA

## Contents

# 1 Estimate of the spin-pumping voltage

Here we estimate amplitude of the artifact voltage generated by spin pumping together with the inverse spin Hall effect in our devices: [1–4]

$$S_{XX}^{\text{art, est}} = V_{\text{sp}} = -\frac{eB_0 R_{\text{tot}} \xi^{\text{DL}}_{\parallel}}{2\pi\alpha^2 \gamma T_{\text{int}}(2B_0 + \mu_0 M_{\text{eff}})^2} g_{\text{eff}}^{\uparrow\downarrow} \lambda_{\text{sd}} \tanh\left(\frac{t_{\text{ZrTe}}}{2\lambda_{\text{sd}}}\right)$$
$$\times \left[(\tau_\parallel^0)^2 + \left(1 + \frac{\mu_0 M_{\text{eff}}}{B_0}\right)(\tau_\perp^0)^2\right]$$
$$\times S(B)\cos^2\phi \begin{cases} L \sin\phi & \text{longitudinal} \\ W \cos\phi & \text{transverse} \end{cases} \quad (1)$$

Here $R_{\text{tot}}$ is the total device resistance, $T_{\text{int}}$ is an interface transparency for spin currents going from the ZrTe$_3$ to the Py (for estimates, we assume $T_{\text{int}} \approx 1$), $g_{\text{eff}}^{\uparrow\downarrow}$ is the effective spin mixing conductance, and $\lambda_{\text{sd}}$ is the spin diffusion length in the ZrTe$_3$. In our estimate we use values for $B_0$, $M_{\text{eff}}$, $\alpha$, $\tau_\parallel^0$, $\tau_\perp^0$ and $\xi_\parallel^{\text{DL}}$ determined from the Hall ST-FMR measurements, $R_{\text{tot}}$ from 2-point resistance measurements, and $t_{\text{ZrTe}}$ from atomic force microscopy. We assume a spin-mixing conductance of $g_{\text{eff}}^{\uparrow\downarrow} \approx 2 \times 10^{19}$ m$^{-2}$ as measured for Pt/Py bilayers [3] and a spin diffusion length of $\lambda_{\text{sd}} \approx 2$ nm which is typical for strong spin-orbit materials. [5, 6] **Supplementary Figure 1** shows that the estimated artifact voltage $S_{XX}^{\text{art, est}}$ has the same order of magnitude as the measured signal $S_{XX}^{\text{AMR}}$ for ZrTe$_3$ layers thicker than about 10 nm. This agrees with our conclusions in the main text based on the Hall-detected ST-FMR measurements that the standard ST-FMR analysis is dominated by artifacts in this thickness range, and indicates that spin pumping is a major contributor to the artifact voltage. Given uncertainties in the values of $g_{\text{eff}}^{\uparrow\downarrow}$ and $\lambda_{\text{sd}}$, the Hall analysis presented in the main text (Fig. 8(a)) gives more precise measurements of the artifact voltages.

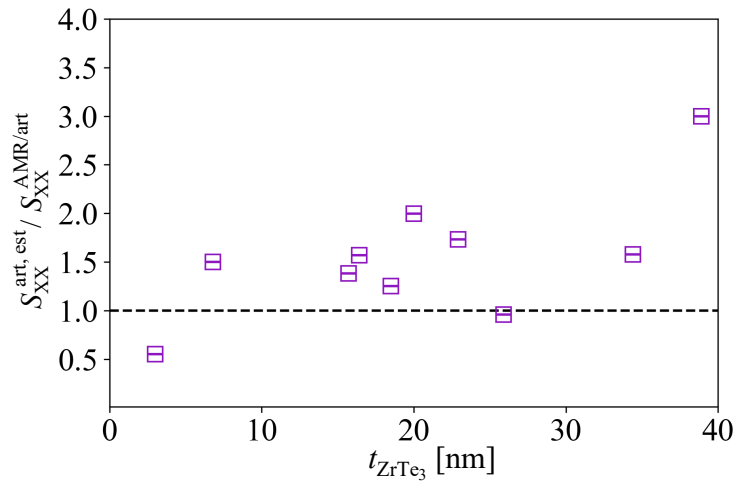

Figure 1: Estimated ratio of spin pumping artifact $S_{XX}^{\text{art,est}}$ to the measured symmetric ST-FMR signal $S_{XX}^{\text{AMR/art}}$ for the longitudinal ST-FMR geometry. The error bars plotted do not include uncertainties in the parameters $g_{\text{eff}}^{\uparrow\downarrow}$ and $\lambda_{\text{sd}}$.

## 2 Hall-ST-FMR calculation of $E_{art}$

The artifact voltage due to spin-pumping can be quantified from the symmetric and antisymmetric resonance amplitudes from the AHE and PHE Hall readouts in addition to the conventional AMR readout. Using Equations (11) and (12) in the main text [4]

$$\frac{-A_{XY}^{AHE}}{S_{XY}^{AHE/art} + W(E_{art}/2)} = \frac{S_{XY}^{PHE/art} + W(E_{art}/2)}{A_{XY}^{PHE}} \quad (2)$$

$$\frac{-A_{XY}^{AHE}}{S_{XY}^{AHE/art} + W(E_{art}/2)} = \frac{S_{XX}^{AMR/art} + L(E_{art}/2)}{A_{XX}^{AMR}} \quad (3)$$

we obtain two independent formulas for $E_{art}$. From Eq. (2):

$$E_{AHE, PHE}^{\pm/art} = -\frac{(S_{XY}^{AHE/art} + S_{XY}^{PHE/art})}{W} \\ \pm \frac{\sqrt{(S_{XY}^{AHE/art} + S_{XY}^{PHE/art})^2 - 4(S_{XY}^{AHE/art} S_{XY}^{PHE/art} + A_{XY}^{AHE} A_{XY}^{PHE})}}{W} \quad (4)$$

and from Eq. (3):

$$E_{AHE, AMR}^{\pm/art} = -\frac{(S_{XY}^{AHE/art} L + S_{XX}^{AMR/art} W)}{WL} \\ \pm \frac{\sqrt{(S_{XY}^{AHE/art} L + S_{XX}^{AMR/art} W)^2 - 4WL(S_{XY}^{AHE/art} S_{XX}^{AMR/art} + A_{XY}^{AHE} A_{XX}^{AMR})}}{WL} \quad (5)$$

Values calculated using either the "+" or "-" sign in Equation (4) can be checked against "+" or "-" values from Equation (5) for consistency. We find that values calculated using $E_{AHE, PHE}^{+/art}$ and $E_{AHE, AMR}^{+/art}$ are consistent. In contrast, the $E_{AHE, PHE}^{-/art}$ values gave an unphysical change in sign even for the thinnest ZrTe$_3$ device (**Supplementary Figure 2**).

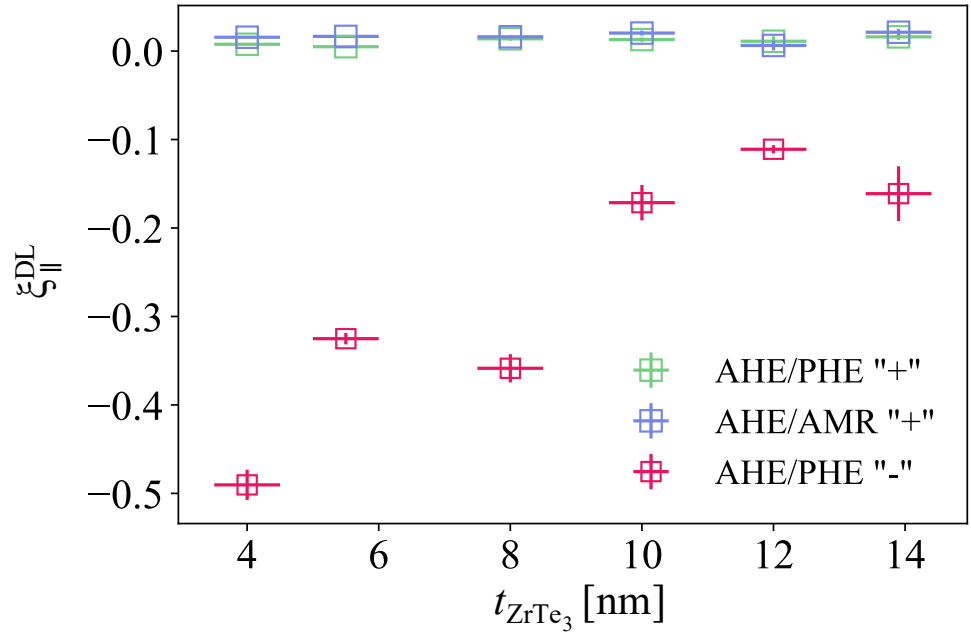

Figure 2: Comparison of corrected $\xi_{||}^{DL}$ values calculated using $E_{AHE,PHE}^{\pm/art}$.

# 3 Raman Spectroscopy

We perform polarized Raman spectroscopy on the fabricated devices and confirm that the ZrTe$_3$ layer retains its as-grown characteristics after the nano-fabrication processes. We use a 532 nm excitation laser on a Witec Alpha 300R confocal Raman microscope with a rotation stage and observe three main peaks at 108, 145 and 213 cm$^{-1}$, in agreement with space group P2$_1$/m as previously reported in the literature.[7, 8]

The relative amplitudes of the peaks change as the angle between the polarization direction and the crystal axis is changed. **Supplementary Figure 3**a,b shows the decrease in the amplitude of the peak at 213 cm$^{-1}$ as the angle $\phi$ between the laser polarisation and device current axis is swept from 0° to 90°. We use this to confirm that the devices were fabricated so that the applied current flows along the b axis perpendicular to the mirror plane of ZrTe$_3$.

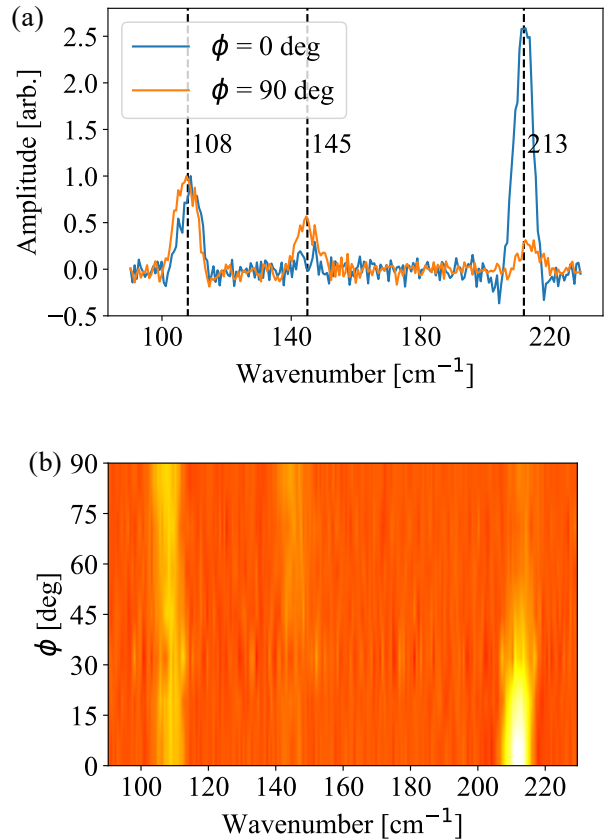

Figure 3: Polarized Raman spectroscopy of a ZrTe$_3$ device used to confirm the crystal quality and crystal-axis direction, where $\phi$ is the angle between the laser polarisation ($E$) and the device current axis ($I$). (a) Comparison between spectra for $\phi = 0°$ ($E \parallel I$) and $\phi = 90°$ ($E \perp I$). (b) Colormap of the spectra evolution as a function of the angle between the polarization and current directions. The Raman peak at 213 cm$^{-1}$ decreases as the polarization angle is rotated from 0° to 90°.

## 4 ZrTe$_3$ resistivity

Assuming that the resistivity of ZrTe$_3$ remains approximately constant with thickness, we can estimate the resistivity of the Permalloy and ZrTe$_3$ layers in our heterostructure using a parallel resistor model

$$\frac{1}{R_{\text{device}}} = \frac{W}{L\rho_{\text{ZrTe}}}t_{\text{ZrTe}} + \frac{W}{L\rho_{\text{Py}}}t_{\text{mag}} \tag{6}$$

where $t_{\text{ZrTe}}$ is the ZrTe$_3$ thickness as measured by atomic force microscopy, $t_{\text{mag}}$ is the thickness of the permalloy layer, and $L$ and $W$ are the device lengths and widths respectively. From a linear fit to L/(WR) vs $t_{\text{ZrTe}_3}$ (**Supplementary Figure 4**) we calculate an average $\rho_{\text{ZrTe}_3}$ = 570 ± 70 μΩcm. We note that ZrTe$_3$ has been reported to crystallize into two polymorths with resistivities ranging from 300 to 700 μΩcm.[7] This suggests that our ZrTe$_3$ flakes are likely a mixture of the two phases. We also obtain $\rho_{\text{Py}}$ = 90 ± 10 μΩcm, similar to previous values reported in our group for glancing angle sputtered Py.[9, 10]

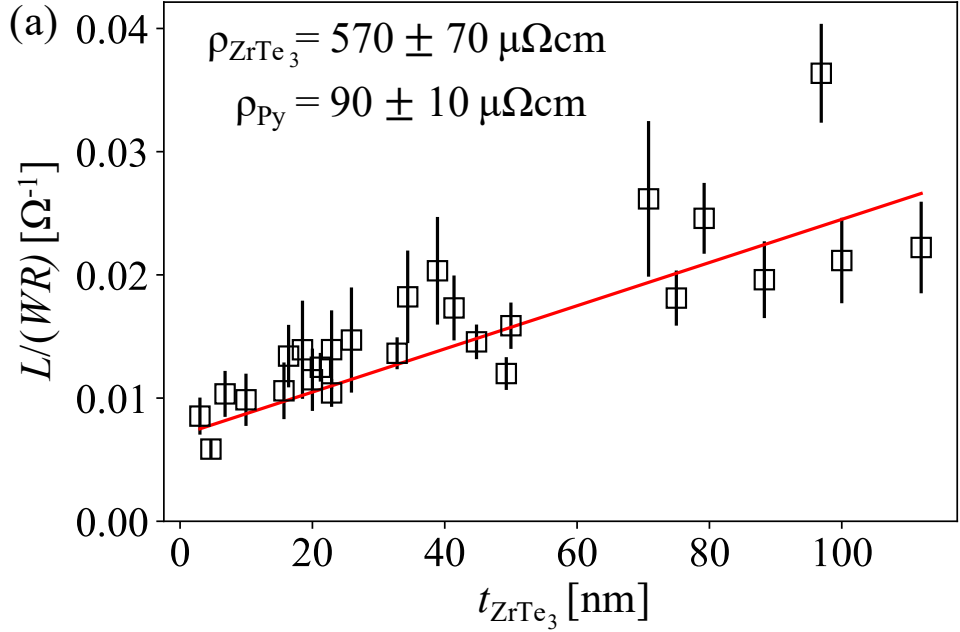

Figure 4: Linear fit of the device-geometry-normalized conductance $L/(WR)$ versus the thickness of the ZrTe$_3$ layer.

# 5 Spin conductivity from $I_{RF}$ calibration

To calibrate $I_{RF}$, we use an Agilent 8722ES 40 GHz vector network analyzer (VNA) to measure the reflection coefficient ($S_{11}$) of each device and transmission coefficient of the RF circuit in the absence of the device ($S_{21}$). The VNA was calibrated using open, short and 50-ohm standards prior to the measurements. We calculate $I_{RF}$ using the $S_{11}$ and $S_{21}$ values at 9 GHz:

$$I_{\text{RF}} = 2\sqrt{\frac{1mW \cdot 10^{\frac{P(dBm)+S_{21}}{10}}(1-|\Gamma^2|)}{50\,\Omega}} \qquad (7)$$

where $P$ is the power of the RF microwave source in dBm, $S_{21}$ is negative, and $\Gamma$ is calculated from the reflection coefficients:

$$\Gamma = 10^{S_{11}(dBm)/20}. \qquad (8)$$

$I_{RF}$ values for each device can be found in Supplementary section 5. We use the standard deviation of the circuit dependent $S_{21}$ transmission values upon repeated measurements to estimate the uncertainties for $I_{RF}$. The spin conductivities are then calculated using:

$$\sigma_{S,A} = -\frac{2\alpha M_s t_{\text{mag}} L(2B_0 + \mu_0 M_{eff})}{I_{\text{RF}}^2 \cdot 50\,\Omega}\left(\frac{1-\Gamma}{1+\Gamma}\right)\left(\frac{dR}{d\phi}\right)^{-1} V_{S,A} \qquad (9)$$

where $\alpha$ is the Gilbert damping, $M_s$ is the saturation magnetization, $t_{\text{mag}}$ is the thickness of the magnetic layer, and $L$ is the device length. We obtain $V_{S,A}$ from the ST-FMR measurements and $\frac{dR}{d\phi}$ from magnetoresistance measurements as a function of in-plane magnetic field angle. We can compare the field-like spin conductivity to the Oersted field contribution calculated using the equation $\sigma_{\text{Oe}} = \frac{e}{\hbar}\mu_0 M_{eff} t_{\text{mag}} t_{\text{ZrTe}_3} \sigma_{\text{ZrTe}_3}$, where $\sigma_{\text{ZrTe}_3} = 1/\rho_{\text{ZrTe}_3}$ (**Supplementary Figure 5**). The uncertainty in the Oersted torque values was estimated from the linear fits for $\rho_{\text{ZrTe}_3}$.

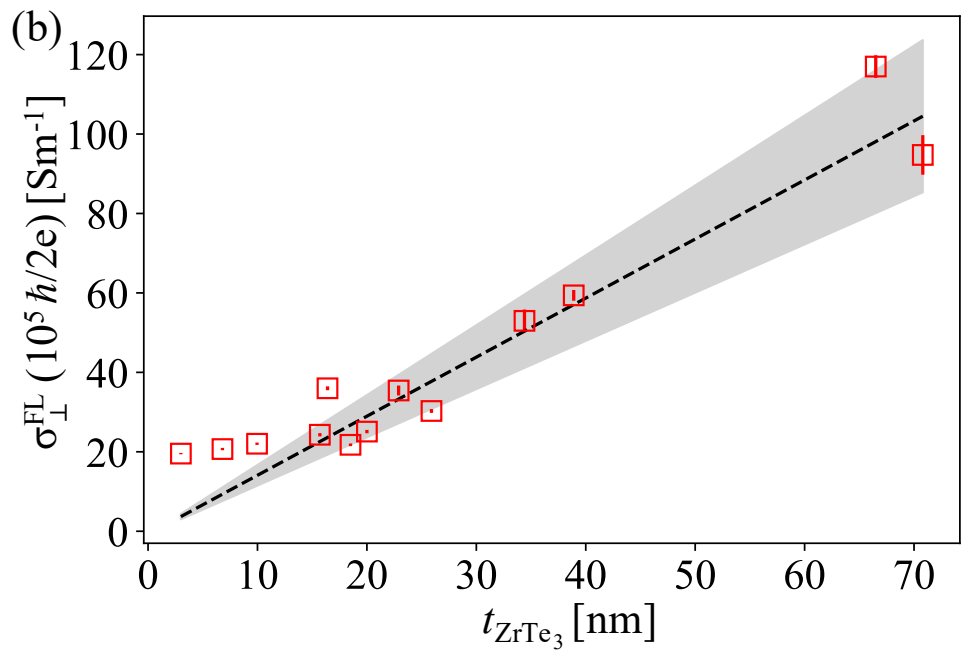

Figure 5: (Field-like spin conductivity as a function of ZrTe$_3$ thickness from current-calibrated STFMR measurements. The Oersted torque (gray shaded region) accounts quantitatively for the field-like torques for ZrTe$_3$ thicker than $\approx 10$ nm.

# 6 Device dimensions

| Device | Type | Thickness [nm] | Length [μm] | Width [μm] | Resistance [Ω] | $I_{RF}$ (mA) |
|---|---|---|---|---|---|---|
| 1 | STFMR | 3 | 3 | 3 | 118 | 6.3(1) |
| 2 | STFMR | 4.66 | 4 | 3 | 230 | 3.5(2) |
| 3 | STFMR | 6.8 | 3 | 3 | 97 | 7.2(2) |
| 4 | STFMR | 9.96 | 3 | 3 | 81 | 8.0(2) |
| 5 | STFMR | 15.7 | 2.5 | 2.5 | 92 | 7.9(2) |
| 6 | STFMR | 16.4 | 3 | 3 | 75 | 8.4(2) |
| 7 | STFMR | 18.5 | 2 | 2 | 80 | 8.1(2) |
| 8 | STFMR | 20 | 2.5 | 2.5 | 87.3 | 7.7(2) |
| 9 | STFMR | 22.9 | 2.5 | 2.5 | 72 | 8.6(2) |
| 10 | STFMR | 25.9 | 2 | 2 | 66 | 9.1(2) |
| 11 | STFMR | 34.4 | 3 | 3 | 55 | 10.1(2) |
| 12 | STFMR | 38.9 | 3 | 3 | 50 | 10.7(2) |
| 13 | STFMR | 57.1 | 13 | 5 | 72 | 9.1(2) |
| 14 | STFMR | 66.5 | 13 | 4 | 72 | 8.5(2) |
| 15 | STFMR | 70.8 | 3 | 3 | 37.2 | 12.6(2) |
| 16 | Hall STFMR | 4 | 4 | 2.2 | 365 | 2.4(1) |
| 17 | Hall STFMR | 5.5 | 5 | 3 | 221 | 3.9(1) |
| 18 | Hall STFMR | 6.5 | 4 | 2.5 | 250 | 3.5(1) |
| 19 | Hall STFMR | 8 | 6 | 3 | 200 | 4.2(1) |
| 20 | Hall STFMR | 10 | 5 | 4 | 39 | 12.3(2) |
| 21 | Hall STFMR | 12 | 4 | 2 | 250 | 3.5(1) |
| 22 | Hall STFMR | 13.9 | 5.5 | 3.5 | 180 | 4.6(1) |
| 23 | Hall STFMR | 21.2 | 6 | 4 | 120 | 6.2(1) |
| 24 | Hall STFMR | 22.9 | 5 | 3 | 160 | 5.1(1) |
| 25 | Hall STFMR | 32.8 | 6 | 4 | 110 | 6.6(1) |
| 26 | Hall STFMR | 41.4 | 4 | 3 | 77 | 8.3(2) |
| 27 | Hall STFMR | 44.8 | 6 | 4 | 103 | 6.9(2) |
| 28 | Hall STFMR | 49.3 | 5 | 3 | 139 | 5.6(1) |
| 29 | Hall STFMR | 50 | 5 | 3 | 105 | 6.8(2) |
| 30 | Hall STFMR | 75 | 5 | 3 | 92 | 7.4(2) |
| 31 | Hall STFMR | 79.2 | 6 | 4 | 61 | 9.5(2) |
| 32 | Hall STFMR | 88.3 | 4 | 3 | 68 | 8.9(2) |
| 33 | Hall STFMR | 96.9 | 8 | 4 | 55 | 10.1(2) |
| 34 | Hall STFMR | 100 | 4 | 3 | 63 | 9.4(2) |
| 35 | Hall STFMR | 112 | 4 | 3 | 60 | 9.6(2) |

\clearpage
\end{document}